\documentclass[12pt,british,refpage,intoc,bibliography=totoc,index=totoc,BCOR=7.5mm,captions=tableheading]{extarticle}
\usepackage[T1]{fontenc}
\usepackage{tikz}
\usetikzlibrary{positioning,shapes.geometric, arrows}
\usepackage[latin9]{inputenc}
\usepackage[a4paper]{geometry}
\usepackage{algorithm}
\usepackage{algpseudocode}
\usepackage{color}
\usepackage{babel}
\usepackage{amsmath}
\usepackage{amsfonts}
\usepackage{bbm}
\usepackage{amsthm}
\usepackage{amssymb}
\usepackage{graphicx}
\usepackage{subfig}
\usepackage[numbers]{natbib}
\usepackage[unicode=true,pdfusetitle,
 bookmarks=true,bookmarksnumbered=true,bookmarksopen=false,
 breaklinks=false,pdfborder={0 0 1},backref=false,colorlinks=true]
 {hyperref}
\hypersetup{
 pdfborderstyle=,linkcolor=black,citecolor=blue,urlcolor=black,filecolor=blue,pdfpagelayout=OneColumn,pdfnewwindow=true,pdfstartview=XYZ,plainpages=false}

\makeatletter

\numberwithin{equation}{section}
\numberwithin{figure}{section}
\theoremstyle{plain}

\theoremstyle{remark}

\theoremstyle{remark}
\newtheorem*{rem*}{\protect\remarkname}
\theoremstyle{plain}

\theoremstyle{plain}

\newcommand\restr[2]{\ensuremath{\left.#1\right|_{#2}}}
\newtheorem{Def}{Definition}

\newtheorem{Rmk}[Def]{Remark}

\@ifundefined{date}{}{\date{}}
\usepackage{babel}
\usepackage{caption}
\usepackage[nottoc]{tocbibind}
\allowdisplaybreaks[4]
\usepackage{dsfont}
\DeclareMathAlphabet{\mathcal}{OMS}{cmsy}{m}{n}

\newcommand{\vertiii}[1]{{\left\vert\kern-0.25ex\left\vert\kern-0.25ex\left\vert #1 \right\vert\kern-0.25ex\right\vert\kern-0.25ex\right\vert}}

\providecommand{\lemmaname}{Lemma}

\providecommand{\remarkname}{Remark}
\providecommand{\theoremname}{Theorem}

\makeatother

\providecommand{\corollaryname}{Corollary}
\providecommand{\lemmaname}{Lemma}
\providecommand{\remarkname}{Remark}
\providecommand{\theoremname}{Theorem}

\begin{document}

\title{Implicit Deep Random Vortex Methods (iDRVM) for simulating incompressible flows in wall bounded domains}
\author{By V. Cherepanov\thanks{Mathematical Institute, University of Oxford, Oxford OX2 6GG. Email:
vladislav.cherepanov@maths.ox.ac.uk},$\ \; $S. W. Ertel\thanks{Institut f{\"u}r Mathematik, Technische Universit{\"a}t Berlin, Stra{\ss}e
des 17. Juni 136, 10623 Berlin, DE. Email: \protect\protect\protect\protect\protect\href{mailto:ertel@math.tu-berlin.de}{ertel@math.tu-berlin.de}} }

\maketitle

\begin{abstract}

In this paper we introduce a novel Neural Networks-based approach for approximating solutions to the (2D) incompressible Navier--Stokes equations, which is an extension of so called Deep Random Vortex Methods (DRVM), that does not require the knowledge of the Biot--Savart kernel associated to the computational domain. Our algorithm uses a Neural Network (NN), that approximates the vorticity based on a loss function that uses a computationally efficient formulation of the Random Vortex Dynamics. The neural vorticity estimator is then combined with traditional numerical PDE-solvers, which can be considered as a final implicit linear layer of the network, for the Poisson equation to compute the velocity field. 
The main advantage of our method compared to the standard DRVM and other NN-based numerical algorithms is that it strictly enforces physical properties, such as incompressibility or (no slip) boundary conditions, which might  be hard to guarantee otherwise. The approximation abilities of our algorithm, and its capability for incorporating measurement data, are validated by several numerical experiments.

\medskip{}


\medskip{}

\end{abstract}

    \section{Introduction}

    In recent years, the usage of Neural Networks (NNs) and deep learning techniques for numerical approximations of Partial Differential Equations (PDEs) has become an increasingly popular alternative to classical numerical solvers such as finite difference and finite element schemes. By now there exist many different approaches for NN-based PDE solvers, so that listing all of them is out of the scope of this introduction.
    
    One of the most general NN-based numerical schemes is the Deep Galerkin Method (DGM) \cite{Deep Galerkin}. Similar to Physics-Informed Neural Networks (PINNs) (see e.g. \cite{RaissiPerdikarisKarniadakis}), the DGM directly incorporates the PDE into the loss function, which is usually chosen to be the $L^2$-norm of the PDE residual w.r.t. some probability measure. Boundary conditions  and other properties of the solution of the PDE can be incorporated directly into the loss function by just adding the corresponding (weighted) residual terms. 
    
    A popular alternative are approaches based on stochastic representations of the PDE such as Feynman--Kac formulae \cite{BeckBeckerGrohsJaafariJentzen} or Backward Stochastic Differential equations \cite{HanEJentzen},\cite{BeckEJentzen}. While such methods are restricted to certain classes of second order PDEs, they provide an efficient numerical approximation scheme that can provably overcome the curse of dimensionality \cite{GrohsHornungJentzenvonWurstemberger}.

    The main area of application for NN-based PDE solvers so far seems to be in high dimensional regimes where classical grid based schemes are too computationally expensive and therefore not viable. However even in low dimensional settings accurate approximations of certain PDEs can be prohibitively computationally expensive. In turbulent regimes this is for example the case for the (incompressible) Navier--Stokes equations
	\begin{align}\label{incompressible Navier Stokes}
		\begin{split}
			\partial_t  u
			&=
			\nu \Delta_{x} u
			-\left(u\cdot\nabla_{x}\right) u 
			-\nabla_{x} p
			+
			f
			\quad\text{on}~D
			\\
			\nabla_{x}\cdot u &= 0
                \quad\text{on}~D
                \\
                u&=0\quad\text{on}~\partial D
                ,
		\end{split}
	\end{align}
    which describe the evolution of the velocity field $u$ of some incompressible fluid and  its pressure $p$. The parameter $\nu>0$ denotes the viscosity of the fluid, $f$ some external force and the domain  $D$ is typically two or three dimensional, i.e. $D\subseteq \mathbb{R}^{2}$ or $D\subseteq \mathbb{R}^{3}$. As usual $\Delta_{x}$ and $\nabla_{x}$ denote the (spatial) Laplacian and gradient respectively.

    As direct numerical simulations of \eqref{incompressible Navier Stokes} based on classical solvers can require huge computation hours, NN-based techniques to design efficient approximation schemes have recently received increasing attention. While general NN-based PDE solvers like the aforementioned DGM can in principle also be applied to the Navier--Stokes equations \eqref{incompressible Navier Stokes}, their special structure with the only implicitly determined pressure $p$ makes the design of appropriate loss functions non trivial \cite{Krishnapriyan 2021}.  PINNs for Computational Fluid Dynamics (CFD) have been explored e.g. in \cite{CaiEtAl 2022},\cite{JinEtAl}. Other popular NN-based solvers make use of the specific structure of the Navier--Stokes equations.  One very prominent approach hereby is to use NN-closures for Reynolds-averaged Navier-Stokes, see e.g. \cite{McConkeyYeeLien} and the references found therein.
    
    An alternative probabilistic approach that has recently been proposed in \cite{Deep Random Vortex} is the so called Deep Random Vortex Method (DRVM). It combines a stochastic representation of the Navier--Stokes equations by  Random Vortex Dynamics, with a NN approximation $u^{\theta}$ of the velocity field $u$. The chosen loss function hereby controls the $L^2$-distance between $u^{\theta}$ and $u$.  While this method delivers an efficient approximation scheme, its scope is severely limited by the fact that it requires the explicit knowledge of the  Biot--Savart kernel $K$ associated to the domain $D$. In most applications this kernel will not be available. Also, even though the NN-approximation $u^{\theta}$, might be reasonably close to $u$ in $L^{2}$, it can not be guaranteed that it (even approximately) respects the homogeneous boundary conditions (see also Section \ref{section - velocity vs vorticity}). This is an issue that is shared by many NN-based approaches, as appropriately incorporating such constraints into the NN can be difficult, as it would either require an adaptation of the architecture (final activation layer) or an adaptation of the loss function through Lagrange multipliers. Designing such adaptations efficiently for general computational domains however is a highly challenging task.

    To overcome this problem we therefore derive a novel variant of DRVNs that combines a NN-based approximation of the vorticity with traditional numerical PDE solvers, acting as a final implicit layer of the network, to compute the fluid velocity from a Poisson equation. In each training cycle of the network $\omega^{\theta}$ we first optimize the weights $\theta$ of the NN according to a Monte Carlo approximation $\hat{\mathcal{L}}\left(\theta\right)$ of a loss function $\mathcal{L}\left(\theta\right)$, which is equivalent to the $L^{2}$-distance $\int_{0}^{T}\left\|\omega(\cdot,t)-\omega^{\theta}(\cdot,t)\right\|_{L^{2}(D)}^2\mathrm{d}t$ between the true vorticity $\omega:=\nabla_{x}\wedge u$~and our NN-approximation $\omega^{\theta}$. Similar to the DRVM this loss function is based on a stochastic representation of the Navier--Stokes equations using Random Vortex Dynamics. Then we compute an approximation $u^{\theta}$ of the fluid velocity by (numerically) solving the Poisson equation $-\Delta_{x} u^{\theta}= \nabla_{x}\wedge\omega^{\theta}$, which is a consequence of the incompressibility of the fluid. With this approximate velocity $u^{\theta}$ at hand, we then generate samples from the diffusion process
    \begin{align}\label{SDE - Introduction}
        \mathrm{d}X^{\theta}_{\xi}(t)
        =
        u^{\theta}\left(X^{\theta}_{\xi}(t),t\right)\mathrm{d}t
        +
        \sqrt{2\nu}\mathrm{d}B_t
        ,\quad
        X^{\theta}_{\xi}(0)=\xi,
    \end{align}
    which then in turn lets us compute an updated Monte Carlo approximation $\hat{\mathcal{L}}\left(\theta\right)$ of the loss function, and restart the training cycle (depicted in Flowchart \ref{flowchart - training cycle}, see also Algorithm \ref{NRV Algorithm} for pseudocode).

    %%%%%%%%%%%%%%%%%%%%%%%%%%%%%%%%%%%%%%%%%%%%%%%%%%%%%%%%%%%%%
    %                   tik commands
    %
    \tikzstyle{startstop} = [rectangle, rounded corners, minimum width=3cm, minimum height=1cm,text centered, draw=black]
    \tikzstyle{arrow} = [thick,->,>=stealth]
    %%%%%%%%%%%%%%%%%%%%%%%%%%%%%%%%%%%%%%%%%%%%%%%%%%%%%%%%%%%%

    \begin{figure}
            \centering
            \begin{tikzpicture}[node distance=4.4cm]
        
            \node (train) [startstop] 
                {vorticity approximation $\omega^{\theta}$};
            \node (loss) [startstop, below of=train] 
                {approximate loss $\hat{\mathcal{L}}\left(\theta\right)$};
            \node (solve) [startstop, right of=train,xshift=5cm] 
                {velocity approximation $u^{\theta}$};
            \node (sample) [startstop, below of=solve] 
                {Random Vortex Samples $X^{\theta}$};

            \draw [arrow] (train) -- node[anchor=south] {\small solve~$-\Delta_{x} u^{\theta}= \nabla_{x}\wedge\omega^{\theta}$} (solve);

            \draw [arrow] (solve) -- node[anchor=west] {\small solve~ SDE \eqref{SDE - Introduction}} (sample);

            \draw [arrow] (sample) -- node[anchor=north] {\small Monte Carlo approx.} (loss);

            \draw [arrow] (loss) -- node[anchor=east] {\small find $\underset{\theta}{\mathrm{argmin}}~  \hat{\mathcal{L}}\left(\theta\right)$} (train);
        \end{tikzpicture}
        
            \caption{Training cycle of our NN-based algorithm. To simulate the Navier--Stokes equations, these training cycles are repeated several times until $\omega^{\theta}$ can be expected to sufficiently approximate $\omega$.}
            \label{flowchart - training cycle}
        \end{figure}
    
    Note that this algorithm produces a NN-based numerical solver for the Navier--Stokes equation in which the NN is trained on artificially generated data. However, as discussed in Section \ref{appendix - PINN}, one can easily adapt our method to incorporate (real world) measurement data by adapting the loss function  $\hat{\mathcal{L}}\left(\theta\right)$ and thus create a PINN.\newline

    This approach has several advantages:
    \begin{itemize}
        \item As traditional PDE solvers usually strictly enforce the homogeneous Dirichlet conditions of $u$, this alleviates the previously discussed issue of incorporating the boundary conditions into the training of the network.

        \item It allows the usage of state of the art numerical solvers for elliptic PDEs as a black box for the computation of the velocity from neural vorticity approximations. Therefore simulations in domains $D$ with complex boundary geometry are possible with our approach. In particular, unlike the DRVM of \cite{Deep Random Vortex}, our method does not require the explicit knowledge of the Biot--Savart kernel, which for most domains $D$ will not be available.

        \item Solving the Poisson equation has a regularizing effect and thus one can expect that this approach gives approximations in the more regular $H^{1}(D)$-Sobolev norm .   
        
    \end{itemize}

    In the following sections we describe our algorithm in detail and show its effectiveness for simulating incompressible fluid flows through several numerical experiments. In  Section \ref{section - Vorticity and its random vortex representation} we briefly introduce the so called Random Vortex Dynamics, a stochastic representation of the Navier--Stokes equation, which are the basis for our approximation scheme. In Section \ref{section - velocity vs vorticity} we show how a loss function for neural vorticity approximations can be constructed from these Random Vortex Dynamics. In this section we also compare our approach of approximating the vorticity with direct approximations of the velocity, which were considered in \cite{Deep Random Vortex}. In Section \ref{Numerical algorithm} we specify our proposed algorithm in detail, including a description in pseudo code.  Furthermore in Section \ref{Periodic boundary conditions} and Section \ref{appendix - PINN} we discuss how the proposed algorithm can be adapted easily for periodic boundary conditions and for incorporating (real world) measurement data. Finally, in Section \ref{simulation results}, we show the effectiveness of our method through several numerical experiments.

    \section{Vorticity and its random vortex representation} \label{section - Vorticity and its random vortex representation}

    Just as in \cite{Deep Random Vortex}, a key component of our Neural PDE-solver are stochastic representations of the vorticity of the Navier--Stokes equation. For the sake of (notational) simplicity we shall assume throughout this paper that we are investigating a two dimensional flow, i.e. $D\subseteq\mathbb{R}^2$, however we note that by using recently found 3D Random Vortex Dynamics \cite{Qian},\cite{QianSueliZhang}, our algorithm can be generalized to domains in $\mathbb{R}^{3}$. For two dimensional velocity fields $u$, their vorticity $\omega$ is a scalar field, defined by $\omega:=\nabla_{x}\wedge u:=\partial_{x_1} u^{2}-\partial_{x_2} u^{1}$. Using the Navier--Stokes equations \eqref{incompressible Navier Stokes} and elementary vector calculus, one shows that $\omega$ satisfies the vorticity equation \cite{Majda and Bertozzi 2002}
	\begin{align}\label{vorticity equation}
		\partial_t\omega
		&=
		\nu\Delta_{x}\omega
		- (u\cdot\nabla_{x}) \omega
		+
		g
        \quad
        \text{on}~D
		,
	\end{align}
    which is a one-dimensional Fokker--Planck equation with drift $u$ and inhomogeneity $g:=\nabla_{x}\wedge f$. Note that the boundary conditions of $\omega$ are only known implicitly through the velocity $u$.

    Notice that while $u$ is only determined within the domain $D\subseteq\mathbb{R}^{2}$, due to the boundary conditions $u(x,t)=0$ for $x\in\partial D,t\geq 0$, we can trivially extend $u$ on the full plane by setting $u(x,t)=0$ for any $x\notin D,t\geq 0$. In particular if $u(\cdot,t)\in H^{1}(D)$ is a Sobolev function of order $1$, then it is easy to check that the same is true for the extension on the full plane, i.e. $u(\cdot,t)\in H^{1}\left(\mathbb{R}^2\right)$.

    With this extension we can define the Random Vortex Dynamics $X$, which is the family of diffusion processes, defined by the stochastic differential equations (SDEs)
    \begin{align}\label{vortex dynamics - definition}
        \begin{split}
            \mathrm{d}X_{\eta;\tau}\left(t\right)
        &=        u\left(X_{\eta;\tau}\left(t\right),t\right)\mathrm{d}t
        +
        \sqrt{2\nu}\mathrm{d}B_t
        \quad\text{for}~t\geq\tau,
        \\
        X_{\eta;\tau}\left(\tau\right)
        &=\eta,
        \end{split}        
    \end{align}
    for some standard 2D Brownian motion $\left(B_t\right)_{t\geq 0}$. In the following we will often leave out the starting time parameter $\tau$, when $\tau=0$, i.e. we set $X_{\eta}:=X_{\eta;0}$ for any $\eta\in\mathbb{R}^2$.
    
    Their transition probability densities  
    \begin{align}\label{transiton kernel - definition}
        p_{u}\left(\tau,\eta,t,x\right)
        :=
        \mathbb{P}\left( X_{\eta;\tau}\left(t\right)\in\mathrm{d}x \right)
        ,
    \end{align}
    are fundamental solutions of the vorticity equation \eqref{vorticity equation}. But only on the full plane $D=\mathbb{R}^2$ they also coincide with the Green's function (i.e. the integral kernel through which the solutions to \eqref{vorticity equation} can be represented via convolution), due to the absence of boundary conditions.

    With these definitions in mind one can derive a representation of the vorticity $\omega$ by the family of diffusion processes $X$. To simplify the exposition, we assume that the domain is the upper-half plane $\{x \in \mathbb{R}^2: x_2 > 0\}$, however the representations presented below are valid for more general domains (see \cite{CherepanovQian}). 

    Let us denote by $\zeta$ the values of the vorticity at the boundary, i.e. $\zeta(x_1, t) := \omega((x_1, 0), t)$ for $x_1 \in \mathbb{R}$. Then we introduce a perturbation of the vorticity as follows. For a smooth cut-off function $\varphi: [0,\infty) \to [0,1]$, such that $\varphi(r)=1$ for $r\in[0,1/2]$ and $\varphi(r)=0$ for $r\geq 1$, define the perturbation $\sigma_{\varepsilon}(x,t):=\varphi(x_2/\varepsilon)\zeta(x_1,t)$ for some $\varepsilon > 0$, and let
    \begin{equation}
    \omega_{\varepsilon}(x,t)=\omega(x,t)-\sigma_{\varepsilon}(x,t).
    \end{equation}
    Then 
    \begin{equation*}
    \partial_t \omega_{\varepsilon} = \nu\Delta_{x}\omega_{\varepsilon} - (u\cdot\nabla_{x}) \omega_{\varepsilon} + g_{\varepsilon}, 
    \end{equation*}
    with $\left.\omega_{\varepsilon}\right|_{\partial D}=0$, and the perturbed external vorticity $g_{\varepsilon}$ is given by
    \begin{equation}
    g_{\varepsilon}=g-\partial_t \sigma_{\varepsilon}-(u\cdot\nabla_{x})\sigma_{\varepsilon}+\nu\Delta_{x}\sigma_{\varepsilon}.
    \end{equation}
    As shown in \cite{CherepanovQian}, the perturbed vorticity satisfies the following representation
    \begin{align}
    \omega_{\varepsilon}(\xi,t)&=\int_{D} \mathbb{E}^{\eta\rightarrow\xi}\left[1_{\{t<\tau_t(X_{\eta})\}}\right] \omega_{\varepsilon}(\eta,0) p_u(0,\eta,t,\xi)\textrm{d}\eta \nonumber\\    &+\int_{0}^{t}\int_{D}\mathbb{E}^{\eta\rightarrow\xi}\left[1_{\{t-s<\tau_t(X_{\eta})\}}g_{\varepsilon}(X_{\eta}(s),s)\right]p_{u}(0,\eta,t,\xi)\textrm{d}\eta\textrm{d}s,
    \end{align}
    where $\tau_t$ is the last hitting time of the boundary $\partial D$ before time $t$, and $\mathbb{E}^{\eta\rightarrow\xi}$ denotes the expectation conditioned on $X_{\eta}(t)=\xi$. Then, given any (appropriate) test function $\phi$, this provides the weak formulation of the vorticity in the limit as $\varepsilon \to 0$
    \begin{align}\label{stochastic representation of vorticity}
    \begin{split}
    \int_{D}\omega(x,t)\phi(x)\mathrm{d}x &= \int_{D} \mathbb{E}\left[\phi\left(X_{\eta}(t)\right) 1_{\{t<\tau_t(X_{\eta})\}}\right] \omega (\eta, 0) \textrm{d} \eta \\
    &+ \lim_{\varepsilon \to 0} \int_{D} \mathbb{E}\left[\phi\left(X_{\eta}(t)\right) \int_0^t g_{\varepsilon}(X_{\eta}(s), s)1_{\{t-s<\tau_t(X_{\eta})\}} \textrm{d}s \right] \textrm{d} \eta.
    \end{split}
    \end{align}
    In the perturbation of external vorticity $g_{\varepsilon} - g$, the only term that contributes to the limit is $\rho_{\varepsilon}(x,t) := \nu \frac{\partial^2}{\partial x_2^2} \sigma_{\varepsilon}(x,t) = \frac{\nu}{\varepsilon^{2}}\varphi''(x_{2}/\varepsilon)\zeta(x_{1},t)$. In general, its limit is represented as an integral with $\frac{\partial}{\partial n}p^{D}(0,\eta,t,\xi)$, where $\frac{\partial}{\partial n}$ is the normal derivative at the boundary in $\eta$ variables --- one writes it as the following boundary integral
    \begin{equation}
    \nu\int_{0}^{t}\int_{\partial D}\frac{\partial}{\partial n}\mathbb{E}\left[\phi\left(X_{\eta}(t)\right)1_{\{t-s<\tau_t(X_{\eta})\}}\right]\zeta(\eta,s)\textrm{d}\eta\textrm{d}s.
    \end{equation}
    In practice, however, we approximate the term computing for some small $\varepsilon$ the following
    \begin{equation}\label{Omega_definition}
    \Omega_{\varepsilon}(\eta, t) := \omega (\eta, 0) 1_{\{t<\tau_t(X_{\eta})\}} +\int_0^t \left(g(X_{\eta}(s), s)+\rho_{\varepsilon}(X_{\eta}(s),s)\right)1_{\{t-s<\tau_t(X_{\eta})\}} \textrm{d}s
    \end{equation}
    and writing
    \begin{align}\label{testfunction_integral_with_Omega}
    \begin{split}    \int_{D}\omega(x,t)\phi(x)\mathrm{d}x\approx\int_{D}\mathbb{E}\left[\Omega_{\varepsilon}(\eta,t)\phi\left(X_{\eta}(t)\right)\right]\mathrm{d}\eta.
    \end{split}
    \end{align}
    In the following, we will often omit the parameter $\varepsilon$ in $\Omega_{\varepsilon}$ to ease the notation. 

    Thus, given samples $X^i,~i=1,\ldots, N$ of the diffusion process \eqref{vortex dynamics - definition}, one can approximate the integration of the vorticity $\omega$ against test functions $\phi$. However to generate such samples one would need to know the velocity $u$ a priori. Instead one uses approximate samples. To this end one first rewrites the velocity $u$ in \eqref{vortex dynamics - definition} as an integral of the vorticity $\omega$, turning \eqref{vortex dynamics - definition} into a closed system of McKean--Vlasov equations. Then one uses a mean-field interacting particle system to generate approximate samples. To write \eqref{vorticity equation} into a closed form, one can use that $u$ is divergence free, to show that
	\begin{align}\label{Poisson equation velocity}
		-\Delta_{x} u 
		= \nabla_{x} \wedge \omega
		:= \left(
			\begin{matrix}
				\partial_{x_2}\omega\\
				-\partial_{x_1}\omega				
			\end{matrix}
		\right)
		\quad\text{on}~D
		.
	\end{align} 
 
    Let now $\mathcal{G}:D\times D\to\mathbb{R}$ be the Green's function of $\left(-\Delta_{x}\right)$ on $D$, then the Biot--Savart kernel $K:D\times D\to\mathbb{R}$ is defined by
     \begin{align}\label{Biot-Savart kernel}
         K(x,y)
         :=
         \nabla_{x} \wedge \mathcal{G}
		:= \left(
			\begin{matrix}
				\partial_{x_2}\mathcal{G}(x,y)\\
				-\partial_{x_1} \mathcal{G}(x,y)				
			\end{matrix}
		\right)
    .
     \end{align}
    Combining the Poisson equation  \eqref{Poisson equation velocity} with the definition of the Biot--Savart kernel implies the so called Biot--Savart law
    \begin{align}\label{Biot-Savart law}
        u(y,t)
        &=\int_{D}K(x,y)\omega(x,t)\mathrm{d}x,
    \end{align}
    which in turn can be combined with the stochastic representation \eqref{stochastic representation of vorticity}, to rewrite the velocity $u$ into
    \begin{align}\label{stochastic representation of velocity}
        \begin{split}
            u(y,t)
    &=\int_{D} \mathbb{E}\left[K\left(X_{\eta}(t),y\right) 1_{\{t<\tau_{t}(X_{\eta})\}}\right] \omega (\eta, 0) \textrm{d} \eta \\
    &+ \lim_{\varepsilon \to 0} \int_{D} \mathbb{E}\left[K\left(X_{\eta}(t),y\right) \int_0^t g_{\varepsilon}(X_{\eta}(s), s)1_{\{t-s<\tau_{t}(X_{\eta})\}} \textrm{d}s \right] \textrm{d} \eta.
    \end{split} 
    \end{align}

    Thus the Random Vortex Dynamics \eqref{vortex dynamics - definition} can be rewritten into closed form as a flow of McKean--Vlasov equations. As usual these McKean--Vlasov equations can be approximated by a system of interacting (ordinary) SDEs, leading to a Monte--Carlo approximation of the velocity, called Random Vortex Methods (RVM). In the simplest setting, namely $D=\mathbb{R}^2, g=0$, this is a classical method in Computational Fluid Dynamics, first investigated by Chorin \cite{Chorin 1973} in the 1970s, see also e.g. \cite{Cottet95},\cite{Goodman1987},\cite{Long1988}. Recently efficient variants of the RVM for some domains involving a boundary and an external force have been derived based on the stochastic representation \eqref{stochastic representation of vorticity} above, see \cite{CherepanovErtelQianWu},\cite{CherepanovQian},\cite{Qian}. 
    
    Nevertheless vanilla Monte Carlo simulations can be inefficient and for general domains designing accurate numerical approximations based on the stochastic representation \eqref{stochastic representation of vorticity} above can be challenging, in particular because the Biot--Savart kernel will usually be unknown. Therefore, similar to the DRVM from \cite{Deep Random Vortex}, we propose a NN-based alternative to the standard Monte Carlo approximation of \eqref{stochastic representation of vorticity} and thus also of $u$. The usage of NNs in our algorithm and its comparison to the previously derived DRVM, is discussed in the next section.

    \begin{Rmk}\label{Remark simplified representation}
        Note that for certain settings formula \eqref{stochastic representation of velocity} can be simpilfied significantly. E.g. if the velocity evolves on the full plane, i.e. $D=\mathbb{R}^2$, and if no external force influences the vorticity, i.e. $g=0$, then \eqref{stochastic representation of velocity} can be simplified into
        \begin{align}\label{velocity formula for full plane}
            u(y,t)
            &=\int_{D} \mathbb{E}\left[K\left(X_{\eta}(t),y\right) \right] \omega (\eta, 0) \textrm{d} \eta,
        \end{align}
        where the Biot--Savart kernel is given by $K(x,y)=\frac{1}{2\pi}\frac{(x-y)^{\perp}}{\|x-y\|^2}$. This was the setting considered by Chorin \cite{Chorin 1973} and other classical works on RVMs.
        Identity \eqref{velocity formula for full plane} also holds true if the domain $D$ is the two dimensional torus, the homogenous Dirichlet boundary conditions are replaced by periodic boundary  conditions and $K$ is the Biot--Savart kernel on the torus.
        However for general domains $D\subseteq \mathbb{R}^2$, the simplified representation formula \eqref{velocity formula for full plane} can not be expected to hold.
    \end{Rmk}
    
    \section{Approximating velocity vs. vorticity}\label{section - velocity vs vorticity}

    The DRVM, proposed by \cite{Deep Random Vortex}, used NNs to approximate the velocity directly. These networks are trained to minimize the $L^2$-distance to the true velocity $u$, which is approximated by using Monte Carlo approximations based on the formula \eqref{velocity formula for full plane} in Section \ref{section - Vorticity and its random vortex representation}. More precisely, consider a NN (or, more generally, other function approximators) $u^{\theta}:\mathbb{R}^{2}\times [0,T]\to\mathbb{R}^2$, where $\theta$ denotes the trainable parameters/weights. To construct the loss function denote by $X^{\theta}$ the family of diffusions\
    \begin{align}\label{neural Random Vortex Dynamics}
        \begin{split}
            \mathrm{d}X^{\theta}_{\eta}\left(t\right)
        &=        u^{\theta}\left(X^{\theta}_{\eta}\left(t\right),t\right)\mathrm{d}t
        +
        \sqrt{2\nu}\mathrm{d}B_t
        \quad\text{for}~t\geq 0,
        \\
        X^{\theta}_{\eta}\left(0\right)
        &=\eta.
        \end{split}        
    \end{align}
    With this notation, the loss function $\mathcal{L}$ used for training the NN was constructed from \eqref{velocity formula for full plane} as
    \begin{align}\label{loss function deep random vortex - velocity}
        \mathcal{L}(\theta)
        :=
        \int_{0}^{T} \int_{D}
            \left\|
                u^{\theta}(y,t)
                -
                \int_{D} \mathbb{E}\left[K\left(X^{\theta}_{\eta}(t),y\right) \right] \omega (\eta, 0) \textrm{d} \eta           
            \right\|^2
        \mathrm{d}y~\mathrm{d}t.
    \end{align}

    \begin{Rmk}
        Note that while in the loss function \eqref{loss function deep random vortex - velocity}, only the evaluations of $u^{\theta}$ inside the domain $D$ appear, the SDE \eqref{neural Random Vortex Dynamics} requires $u^{\theta}$ to be defined on the whole plane $\mathbb{R}^{2}$.
    \end{Rmk}

    As mentioned in Remark \ref{Remark simplified representation}, approximations based on this loss function \eqref{loss function deep random vortex - velocity} can only be expected to result in good approximations of $u$ in certain settings  that do not involve boundaries with no slip conditions, e.g. when the domain $D$ is the full plane $\mathbb{R}^{2}$ or the torus, as more general domains would require adaptations based on the more complex representation formula \eqref{stochastic representation of velocity}. However, even with such adaptations, this approach requires knowledge of the Biot--Savart kernel $K$ associated to the computational domain $D$, which is not known explicitly for most domains $D$.

    Furthermore, even in the few cases where $K$ is known explicitly and the representation formula \eqref{velocity formula for full plane} holds, the minimization of the loss function \eqref{loss function deep random vortex - velocity} can at best be expected to result in an approximation of the true velocity with respect to the $L^{2}$-norm. However  $L^{2}$-convergence can of course not guarantee that the approximations capture the right behaviour of the fluid near the boundary, since the trace $\restr{u^{\theta}}{\partial D\times[0,T]}$ is of course not continuous with respect to the $L^{2}$-norm (see Remark \ref{appendix - discontinuity of trace wrt L2} for a very elementary counterexample). This is in particular problematic for many engineering applications, where one is often primarily interested in the behaviour of the fluid near the boundary in order to compute  boundary stress.

    \begin{Rmk}\label{appendix - discontinuity of trace wrt L2}
        For any function $u:\mathbb{R}^{2}\to\mathbb{R}$ one can easily construct a sequence of perturbations $\left(u^{\epsilon}\right)_{\epsilon>0}$, such that $\left\|u-u^{\epsilon}\right\|_{L^{2}(D)}\xrightarrow{\epsilon\to 0}0$, but $\inf_{x\in\partial D} \left\|u(x)-u^{\epsilon}(x)\right\|\xrightarrow{\epsilon\to 0}+\infty$. To construct such perturbations, denote the $\epsilon$-neighborhood around the boundary $\partial D$ by $\mathrm{U}_{\epsilon}\left(\partial D\right)$, denote its Lebesgue measure by $\mathrm{Leb}\left(\mathrm{U}_{\epsilon}\left(\partial D\right)\right)$ and let $\phi_{\epsilon}:\mathbb{R}^{2}\to\mathbb{R}$ be a smooth function with
        \begin{align*}
            \mathbbm{1}_{\mathrm{U}_{\epsilon/2}
            \left(\partial D\right)}
            \leq
            \phi_{\epsilon}
            \leq
            \mathbbm{1}_{\mathrm{U}_{\epsilon}
            \left(\partial D\right)}.
        \end{align*}
        Then for $u_{\epsilon}(x)
        :=
        u(x)+\frac{\phi_{\epsilon}(x)}{\mathrm{Leb}\left(\mathrm{U}_{\epsilon}\left(\partial D\right)\right)^{1/4}},~x\in\mathbb{R}^{2}$, we have
        \begin{align*}
            \left\|u_{\epsilon}(\cdot,t)-u(\cdot,t)
             \right\|_{L^{2}\left(D\right)}^2
            &\leq
            \mathrm{Leb}\left(\mathrm{U}_{\epsilon}\left(\partial D\right)\right)^{1/2}
            \xrightarrow{\epsilon\to 0} 0
            \\
            \inf_{x\in\partial D}
            \left\|u_{\epsilon}(x)-u(x)
             \right\|^2
             &=
             \frac{1}{\mathrm{Leb}\left(\mathrm{U}_{\epsilon}\left(\partial D\right)\right)^{1/2}}
             \xrightarrow{\epsilon\to 0} +\infty.
        \end{align*}
    \end{Rmk}

    One simple way to address this issue would be an incorporation of the boundary data into the loss function via Lagrange multipliers, similar to what is done in DGM. However, this may negatively impact the training of the network and may result in a worse performance. Another alternative would be to adapt the final (activation) layer of the network to enforce that the boundary conditions are satisified, however for general domains such adapted activation functions may be difficult to find and may also negatively impact the performance of the network. 

    A simple alternative, which we investigate in this paper, is that a NN $\omega^{\theta}$ is only trained to approximate the vorticity $\omega$. The velocity is then approximated by the output $u^{\theta}$ of standard elliptic PDE solvers via the Poisson equation \eqref{Poisson equation velocity}. Note that the continuity of the Poisson equation \eqref{Poisson equation velocity} with respect to the right hand side\footnote{A standard result in PDE theory, see e.g. \cite{Evans}.} implies
    \begin{align}\label{continuity of poisson - inequality 2}
         \left\| u(\cdot,t)- u^{\theta}(\cdot,t)\right\|_{H^{1}(D)}
         \leq
         \sqrt{1+C_{\mathrm{P}}^{2}}
         ~
         \left\| 
            \omega(\cdot,t)-\omega^{\theta}(\cdot,t)
        \right\|_{L^{2}(D)}
    \end{align}
    where $C_{\mathrm{P}}$ is the Poincar\'e constant of the domain $D$.
    Therefore, assuming that  at any time $t\geq 0$ the NN $\omega^{\theta}(\cdot,t)$ approximates the true vorticity $\omega(\cdot,t)$ in $L^{2}(D)$, as well as that the chosen numerical PDE solver returns accurate approximations w.r.t. the $H^{1}(D)$-norm, one can thus expect that with this approach $u^{\theta}$ approximates $u$ in $H^{1}(D)$. In particular, since the restriction to the boundary
    \begin{align*}
        \restr{\cdot}{\partial D}:
            H^{1}(D)
            \to
            L^{2}(\partial D),
    \end{align*}
    is a bounded linear operator on $H^{1}(D)$, one would thus expect that the approximation captures the correct behaviour of the true velocity field $u$ near the boundary.\newline

    To derive a loss function for our NN $\omega^{\theta}$, we use the stochastic representation \eqref{stochastic representation of vorticity} of the vorticity $\omega$. Note that \eqref{stochastic representation of vorticity} only gives us a weak representation in the sense that it only describes the integration of $\omega$ against testfunctions. However, when using gradient-based optimization algorithms, this is enough for training the NN $\omega^{\theta}$ to minimize an $L^{2}$-loss, as by completing the square, we have
    \begin{align*}
        &\left\|\omega^{\theta}-\omega\right\|_{D\times [0,T]}^2
        =
        \int_{0}^{T} \int_{D} \left|\omega^{\theta}(x,t)-\omega(x,t)\right|^2 \mathrm{d}x~\mathrm{d}t
        \\&=
        \int_{0}^{T} \int_{D} \left|\omega^{\theta}(x,t)\right|^2
        \mathrm{d}x~\mathrm{d}t
        -
        2
        \int_{0}^{T} \int_{D} \omega^{\theta}(x,t)~\omega(x,t)
        \mathrm{d}x~\mathrm{d}t
        +
        \int_{0}^{T} \int_{D} \left|\omega(x,t)\right|^2
        \mathrm{d}x~\mathrm{d}t.
    \end{align*}

    Note that the last integral is constant w.r.t. the parameters/weights $\theta$. Thus, gradient descent algorithms (w.r.t. the parameters $\theta$) applied to
    \begin{align}\label{l2 loss without constants}
        \int_{0}^{T} \int_{D} \left|\omega^{\theta}(x,t)\right|^2
        \mathrm{d}x~\mathrm{d}t
        -
        2
        \int_{0}^{T} \int_{D} \omega^{\theta}(x,t)~\omega(x,t)
        \mathrm{d}x~\mathrm{d}t,
    \end{align}
    also minimize the $L^2$-loss $\left\|\omega^{\theta}-\omega\right\|_{D\times [0,T]}^2$.\newline

    Using the approximate form \eqref{testfunction_integral_with_Omega} for the stochastic representation \eqref{stochastic representation of vorticity}, we can rewrite \eqref{l2 loss without constants} into
    \begin{equation}\label{l2 loss without constants - stochastic rep}
    \int_{0}^{T} \int_{D} \left|\omega^{\theta}(\eta,t)\right|^2\mathrm{d}\eta~\mathrm{d}t
    -2\int_{0}^{T} \int_{D} \mathbb{E}\left[\Omega(\eta,t) \omega^{\theta}(X_{\eta}(t),t)\right] \textrm{d}\eta~\mathrm{d}t
    \end{equation}

    However the diffusion process $X$ is unknown, so computing \eqref{l2 loss without constants - stochastic rep} exactly is not possible. Instead we use the Poisson equation \eqref{Poisson equation velocity} to define a velocity approximation $u^{\theta}$, based on the NN $\omega^{\theta}$. Thus define $u^{\theta}$ to be the unique solution of the Poisson equation with homogeneous Dirichlet boundary conditions
    \begin{align}\label{Poisson equation for NN}
        \begin{split}
            -\Delta_{x} u^{\theta}
		  &= \nabla_{x} \wedge    \omega^{\theta}\quad\text{on}~D
            \\
            u^{\theta}&= 0\quad\text{on}~\partial D.
        \end{split}
    \end{align}
    
    Outside the domain $D$ set $u^{\theta}(x,t)=0,~x\notin D,~t\geq 0$ and define the diffusion processes $X^{\theta}$ just as in \eqref{neural Random Vortex Dynamics}. This in turn lets us define the loss function $\mathcal{L}$ by
    \begin{equation}\label{definition loss function - general case}
        \mathcal{L}(\theta)
        :=
        \int_{0}^{T} \int_{D} \left|\omega^{\theta}(\eta,t)\right|^2
        \mathrm{d}\eta~\mathrm{d}t
        -2\int_{0}^{T} \int_{D}\mathbb{E}\left[\Omega(\eta,t)\omega^{\theta}\left(X^{\theta}_{\eta}(t)\right)\right]\textrm{d}\eta~\mathrm{d}t.
    \end{equation}

    \begin{Rmk}
        Note that while only $\omega^{\theta}$ is a traditional (deep) NN and $u^{\theta}$ is obtained through the Poisson equation \eqref{Poisson equation for NN}, one can also view $u^{\theta}$ as a NN with a final convolutional layer given by a convolution with the Biot--Savart kernel $u^{\theta}=K\circledast\omega^{\theta}$. However as the kernel $K$ is usually not explicitly known one should thus rather view $u^{\theta}$ as a NN with a final linear, implicit layer, which in our case is determined by the Poisson equation \eqref{Poisson equation for NN}. Such implicit layers have recently received increasing attention in the Machine Learning community, e.g. as part of Deep Equilibrium Models \cite{BaiKolterKoltun}, where they are usually not given through differential equations, but rather by root finding problems of the form $F\left(u^{\theta},\omega^{\theta}\right)=0$.
    \end{Rmk}

    \section{Numerical algorithm}\label{Numerical algorithm}

    With the general loss function given in \eqref{definition loss function - general case}, we can describe the approximate loss $\hat{\mathcal{L}}$ that is used in simulations by first discretising the time and space integrals
    \begin{equation}\label{discretized loss on full time interval}
    \hat{\mathcal{L}}\left(\theta\right)
    \approx \sum_{k=0}^{N_{\text{steps}}} \left(\sum_{i} 
    \omega^{\theta}(\eta^i,t_k)^2 \Delta\eta^i-2\sum_{i} \mathbb{E}\left[\Omega(\eta^i, t_k) \omega^{\theta}(X_{\eta^i}(t_k),t_k)\right]\Delta\eta^i\right) \Delta t_{k},
    \end{equation}
    and since in practice we compute the approximation $\omega^{\theta}$ at each fixed time $t_k$, it is sufficient to consider
    \begin{equation}\label{discretized_loss}
    \hat{\mathcal{L}}(\theta, t_k):=
    \sum_{i}\left(\omega^{\theta}(\eta^i,t_k)^2-2   \mathbb{E}\left[\Omega(\eta^i, t_k) \omega^{\theta}\left(X_{\eta^i}(t_k),t_k\right)\right]\right)\Delta\eta^i
    \end{equation}
    for the loss function at each step. In the above discretization, we used a finite set of lattice points $\eta^i \in D,~ i = 1, \ldots, N$ with corresponding mesh sizes $\Delta \eta^i$ at each point. 
    % Below we consider two domains, namely half-plane and periodic channel, for which we take the rectangular lattices and the time steps are taken uniform, i.e. $\Delta t_k = T / N_{\text{steps}}$ for all $k$. 

    Notice also that the discretized loss \eqref{discretized_loss} has expectations with respect to stochastic processes $X$. One way to approximate them is to simulate independent copies of $X$ and replace the expectations with averages. Another common way in Random Vortex methods is to run the processes $X$ with independent Brownian motions corresponding to each site $\eta^i$ and omit the expectations. In the algorithm we describe below, the second approach is adopted.
	
    With the loss function we discussed above, we can state the following algorithm for our system. For lattice points $\eta^i \in D,~ i = 1, \ldots, N$, initialize $X^i(0) = X_{\eta^i}(0)$ and $\Omega^i(0) = \Omega(\eta^i, 0)$. Then the system is updated with the following typical step:
    \begin{enumerate}
    \item With the current values of $\Omega^i(t)$ and the processes $X^i(t)$, find the minimizer $\hat{\theta}$ of $\hat{\mathcal{L}}(\cdot, t)$ in \eqref{discretized_loss} and set
    \begin{align*}
    \hat{\omega}(\cdot, t) := \omega^{\hat{\theta}}(\cdot, t),
    \end{align*}
    --- note that we minimize the loss over parameters for a class of NNs, however, any suitable class of function approximators $\mathcal{F}$ can be considered here. 
		
    \item Compute the velocity approximation from \eqref{Poisson equation velocity} with $\hat{\omega}$, i.e. $\hat{u}$ is the numerical solution to
    \begin{equation*}
    \Delta_{x} \hat{u} = -\nabla_{x} \wedge \hat{\omega},
    \end{equation*} 
    which is computed through an applicable numerical solver, e.g. Finite Elements Methods, Finite Difference Schemes, Monte Carlo Methods (like Walk-on-spheres) or, when the kernel is known explicitly, discretisations of the Biot--Savart law \eqref{Biot-Savart law}.
		
    \item Update the processes $X^i$ by
    \begin{align*}
    X^{i}(t+\Delta t)=X^{i}(t)+\hat{u}\left(X^{i}(t),t\right)\Delta t+\sqrt{2\nu}\Delta B^{i}
    \end{align*}
    for $N$ independent Brownian motions $B^{i},~ i=1,\ldots,N$ and update the values of $\Omega^i$ by \eqref{Omega_definition}. 
%\begin{align*}
%\Omega^i(t + \Delta t) = \omega (\eta^i, 0) 1_{\{t<\tau_t(X^i)\}} + \left(S^i(t) + g(X^i(t), t)+\rho_{\varepsilon}(X^i(t),t)\right)1_{\{t-s<\tau_t(X^{\eta})\}} .
%\end{align*}

\item Return to step 1.
\end{enumerate}

A more detailed version of the algorithm above is found below in pseudocode in Algorithm \ref{NRV Algorithm} which is optimized so that the computation complexity at each step is constant in time.

\begin{algorithm}
\caption{Random vortex algorithm.}\label{NRV Algorithm}
\begin{algorithmic}
\State \textbf{Initialise:} $X^i \gets \eta^i$ for some lattice $\eta^i \in D$
\State \textbf{Initialise:} $\text{ind}^i \gets 1$ \Comment{indicators of $X^i$ hitting the boundary}
\State \textbf{Initialise:} $S^i \gets (g(X^i,0)+\rho_{\varepsilon}(X^i,0))\Delta t$ 
\State \textbf{Initialise:} $\Omega^i \gets \omega(X^i,0) + S^i$ 

\For{$k \gets 0$ to $N_{\text{steps}}$}
\State \textbf{Find:} $\hat{\omega}=\omega^{\hat{{\theta}}}(\cdot, t_k),~\text{with}~\hat{\theta}\gets{\mathrm{argmin}_{\theta}}~\hat{\mathcal{L}}(\theta)$
\State \textbf{Solve:} $\Delta_{x} \hat{u} = -\nabla_{x} \wedge \hat{\omega}$ 
\State \textbf{Compute:} $X_{\text{upd}}^i \gets \Hat{u}(X^i, t_k) \Delta t + \sqrt{2\nu} \Delta B^i$
\State \textbf{Update:} $X^i \gets X^i + X_{\text{upd}}^i \cdot \text{ind}^i$
    \If{$X^i$ crosses $\partial D$ at $x^i$}
        \State \textbf{Set:} $X^i \gets x^i$, $\text{ind}^i \gets 0$, $S^i \gets 0$
    \EndIf
\State \textbf{Update:} $S^i \gets S^i + (g(X^i,t_k)+\rho_{\varepsilon}(X^i,t_k))\Delta t$
\State \textbf{Update:} $\Omega^i = \omega(X^i,0) \cdot \text{ind}^i + S^i$
\EndFor
\end{algorithmic}
\end{algorithm}

    \section{Periodic boundary conditions}\label{Periodic boundary conditions}
    
    While no slip conditions usually describe the physically correct behaviour of a fluid at the boundary, also considering other boundary conditions allows for more flexibility when designing numerical experiments. In particular periodic boundaries are of great relevance for benchmarking, because they allow to consider flows in compact domains without restricting the motion of the fluid. In this case the computational domain $D$ is often chosen to be a suitable subset of a rectangle $[0,L_{x}]\times[0,L_{y}]$, that may for example be of the form
    \begin{align*}
        D = [0,L_{x}]\times[0,L_{y}] \setminus O,
    \end{align*}
    for some $O\subseteq (0,L_{x})\times(0,L_{y}) $ representing an obstacle in the fluid flow. Periodicty of the flow may be enforced \begin{enumerate}
        \item\label{horizontal periodicty} only on the left side $\{0\}\times[0,L_{y}]$ and right side$\{L_{x}\}\times[0,L_{y}]$, i.e. $u\left((0,y)^{\mathrm{T}},t\right)=u\left((L_{x},y),t\right)$ for $y\in[0,L_{y}]$

        \item\label{vertical periodicty} only on the bottom $[0,L_{x}]\times\{0\}$ and the top $[0,L_{x}]\times\{L_{y}\}$, i.e. $u\left((x,0)^{\mathrm{T}},t\right)=u\left((x,L_{y}),t\right)$ for $x\in[0,L_{x}]$

        \item\label{full periodicty} on all sides $\partial\left([0,L_{x}]\times[0,L_{y}]\right)$ of the torus, i.e. the conditions of both  previous points \ref{horizontal periodicty} and \ref{vertical periodicty} are satisfied.
    \end{enumerate}
    
    On the other parts of the boundary $\partial D$, in particular on the boundary of the obstacle $\partial O$, one might then consider the usual no-slip conditions. In all three cases the stochastic representation formulas of Section \ref{section - Vorticity and its random vortex representation} can be adapted rather easily for the periodic conditions by considering the random vortex dynamics \eqref{vortex dynamics - definition} modulo the periodicity. In case \ref{full periodicty}, i.e. for  of the fully periodic conditions of all sides of the box $\partial\left([0,L_{x}]\times[0,L_{y}]\right)$,  one considers the stochastic process $\left(X_{\eta}(t) \mod (L_x,L_y)^{\mathrm{T}}\right)_{t\geq0}$ for $X$ defined as in \eqref{vortex dynamics - definition}\footnote{This is equivalent to the periodic processing considered in \cite[Equation (26)]{Deep Random Vortex}}. Extensions for periodicity conditions in only one direction follow similarly.
    
    \begin{Rmk}\label{Period_Rem}
        As just described our Algorithm \ref{NRV Algorithm} can be extended easily to incorporate periodic boundary data. Our numerical experiments, described in Section \ref{simulation results}, verify that it produces correct results without any adaptations to the loss function or the Neural Network architecture. However, while designing classes of Neural Networks that always satisfy given no-slip conditions is hard, one can easily design such architectures for periodic conditions by introducing a first periodic activation layer (periodic preprocessing).
    \end{Rmk}

    \section{Incorporating measurement data/Data Assimilation}\label{appendix - PINN}

    In recent years, partially due to the increasing availability of data, incorporating (real world) measurement data into numerical algorithms is an increasingly popular topic in scientific computing. Especially in numerical weather prediction and adjacent fields such data driven approaches, usually referred to as Data Assimilation (see e.g. \cite{EvensenVossepoelVanLeeuwen}), are often applied as they are capable of significantly improving the forecasting capabilities of numerical models. One popular approach for combining models with data is to minimize a cost function (hence called variational Data Assimilation) that incorporates both the misfit w.r.t. the model and the observation data.\newline

    Since our proposed Algorithm \ref{NRV Algorithm} is based on loss minimization, just as any other PINN, one can easily incorporate measurement data by  simply adapting the loss function. To this end an additional term $\mathcal{R}\left({\theta}\right)$, that measures data misfit, is added, i.e. instead of minimizing $\hat{\mathcal{L}}\left(\theta\right)$ (see \eqref{discretized loss on full time interval}), one would minimize the adapted (empirical) loss function
    \begin{align*}
        \hat{\mathfrak{L}}(\theta)
        :=
        \hat{\mathcal{L}}\left(\theta\right)
        +
        \mathcal{R}\left({\theta}\right),
    \end{align*}
    for training the network $\omega^{\theta}$. The choice of $\mathcal{R}$ will highly depend on the nature of the available data. If one is given actual measurements $\omega_k,~k=1,\ldots, K$ of the vorticity $\omega(x_k,t_k)$ at specific (spacetime) points $t_k\geq 0, x_k\in D$, then the most natural choice of $\mathcal{R}$ is the simple $L^{2}$-loss given by
    \begin{align}\label{L2_loss_vorticity}
        \mathcal{R}\left(\theta\right)
        =
        \frac{1}{K}\sum_{k=1}^{K}
        \left\|\omega^{\theta}(x_k,t_k)-\omega_k\right\|^2.
    \end{align}

    However if one has only the fluid velocity $u$ measurements and not its vorticity, this approach can still be adapted. In this case, when one is given velocity measurements $u_k,~k=1,\ldots, K$ at specific spacetime points $t_k\geq 0, x_k\in D$, then a natural choice of $\mathcal{R}$ is given by
    \begin{align*}
        \mathcal{R}\left(\theta\right)
        =
        \frac{1}{K}\sum_{k=1}^{K}
        \left\|u^{\theta}(x_k,t_k)-u_k\right\|^2,
    \end{align*}
    where $u^{\theta}$ is the velocity approximation determined by the Poisson equation \eqref{Poisson equation for NN}. While at first glance this may seem problematic for training the network, as the velocity approximation is only implicitly known through solving a PDE, the linearity of the Poisson equation \eqref{Poisson equation for NN} lets us compute gradients of $u^{\theta}$, w.r.t. the weights/parameters $\theta$ in a simple manner. Let $\partial_{\theta_i}$ denote the derivative with respect to the $i$-th component of $\theta$, then we note that $\partial_{\theta_i}u^{\theta}$ solves
    \begin{align}\label{Poisson equation for weight derivatives}
        \begin{split}
            -\Delta_{x} \partial_{\theta_i}u^{\theta}
		  &= \nabla_{x} \wedge    \partial_{\theta_i}\omega^{\theta}\quad\text{on}~D
            \\
            \partial_{\theta_i}u^{\theta}&= 0\quad\text{on}~\partial D,
        \end{split}
    \end{align}
    and we can thus compute gradients of $u^{\theta}$ w.r.t. the weights/parameters $\theta$ from the gradients of $\omega^{\theta}$ by solving Poisson equations. In particular in situations where one has access to the Biot--Savart kernel $K$, which is a requirement for the DRVM of \cite{Deep Random Vortex}, one can directly represent $\partial_{\theta_i}u^{\theta}$ through the Biot--Savart law \eqref{Biot-Savart law}
    \begin{align}
        \partial_{\theta_i}u^{\theta}(x,t)
        =
        \int_{D} K(x,y) \partial_{\theta_i}\omega^{\theta}(y,t) \mathrm{d}y.
    \end{align}

    \section{Simulation results}\label{simulation results}

    \textit{Numerical Experiment 1 [Flow with prescribed boundary values]. } In this example, we consider the fully bounded box domain $D=\{(x_1,x_2)\in\mathbb{R}^2: 0< x_1,x_2<1\}$ modelled with the lattice $\Lambda := \{(i_1 h, i_2 h): 0 \leq i_1 \leq N, 0 \leq i_2 \leq N\}$. We take the following explicit solution
    \begin{equation}\label{velocity_NE1}
    u(x_1,x_2) = \left(U_0^2 \frac{x_1}{\nu} e^{-U_0\frac{x_2}{\nu}}, U_0 (e^{-U_0\frac{x_2}{\nu}}-1)\right)
    \end{equation}
    restricted on the domain $D$. This implies that the vorticity $\omega(x_1,x_2,0)=-U_0^3\frac{x_1}{\nu} e^{-U_0\frac{x_2}{\nu}}$ which we use to initialise the system. Notice that as the values of the velocity $u$ are non-zero at the boundary $\partial D$ which implies inhomogeneous boundary conditions in the Poisson equation \eqref{Poisson equation velocity}, which we solve using finite element method discussed below.

    For this experiment, we take the number $N = 50$ which results in 2601 points in total, and the mesh size $h = 1/N = 0.02$; the viscosity constant is taken $\nu = 0.01$ and the initial velocity constant $U_0 = 0.02$. We take the external force being identically zero which results in $g \equiv 0$. Thus, the external vorticity is generated only from the boundary vorticity $\zeta$ contributing to the term $\rho_{\varepsilon}$ in \eqref{Omega_definition}. Here, we choose the boundary cutoff parameter $\varepsilon = 0.1$ with the cutoff function
    \begin{equation*}
        \varphi(r) = 2(r-0.5)^3-\frac{3}{2}(r-0.5)+\frac{1}{2}
    \end{equation*}
    for $r \in [0, 1]$, and zero otherwise. 

\begin{figure}
    \centering
    \subfloat[\centering $t=0.01$]{{\includegraphics[width=.3\linewidth]{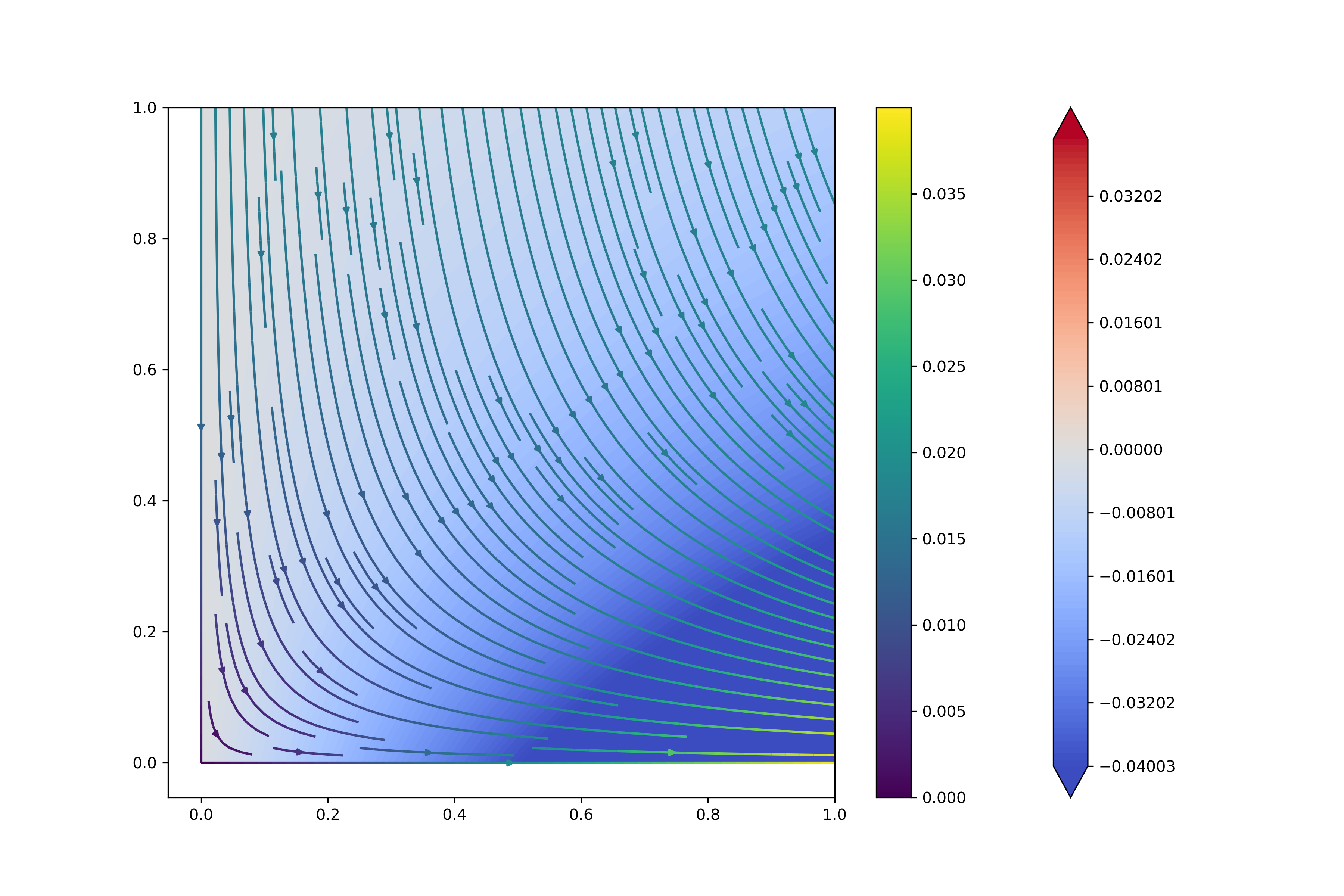} }}
    \subfloat[\centering $t=0.5$]{{\includegraphics[width=.3\linewidth]{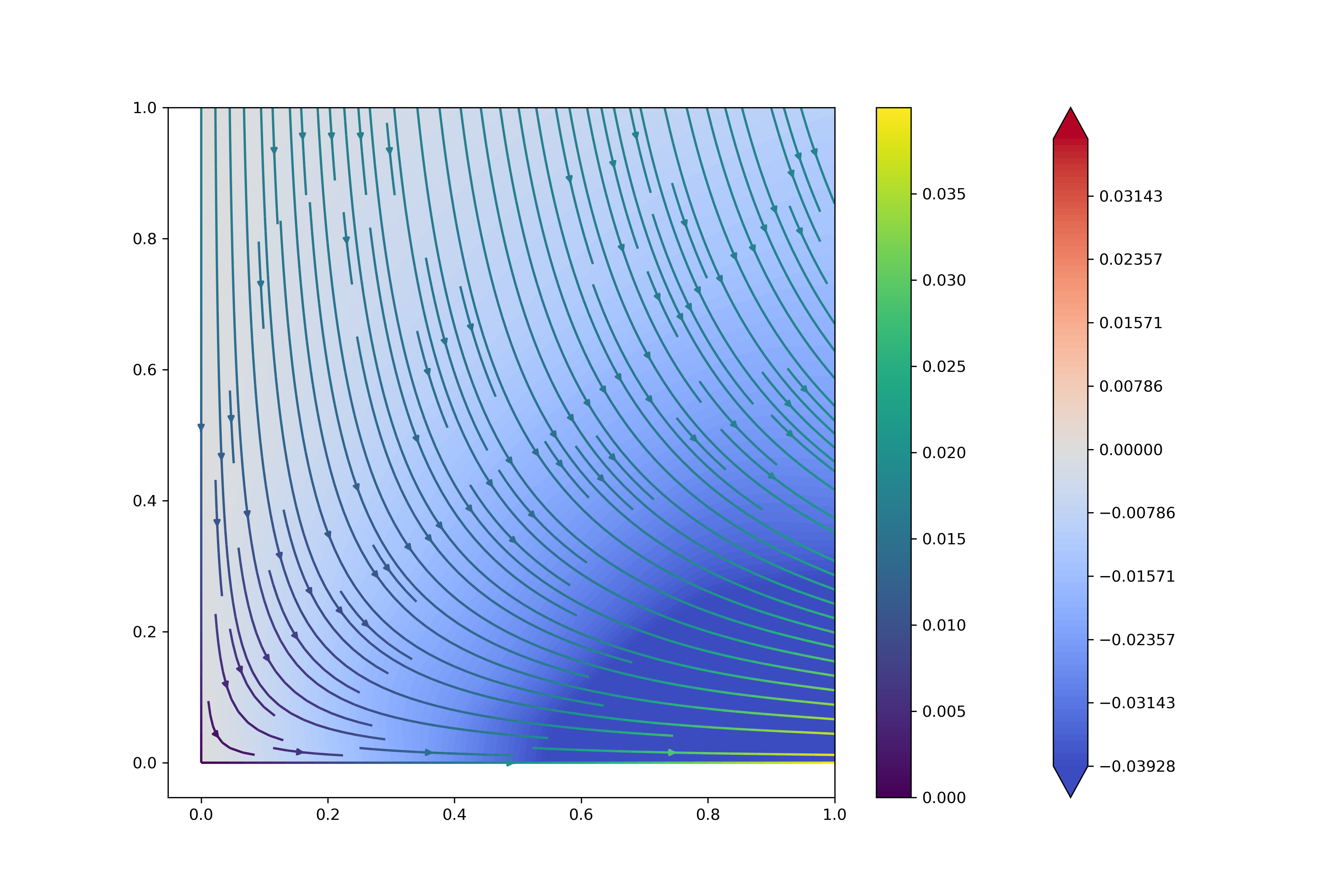}}}
    \subfloat[\centering $t=1.0$]{{\includegraphics[width=.3\linewidth]{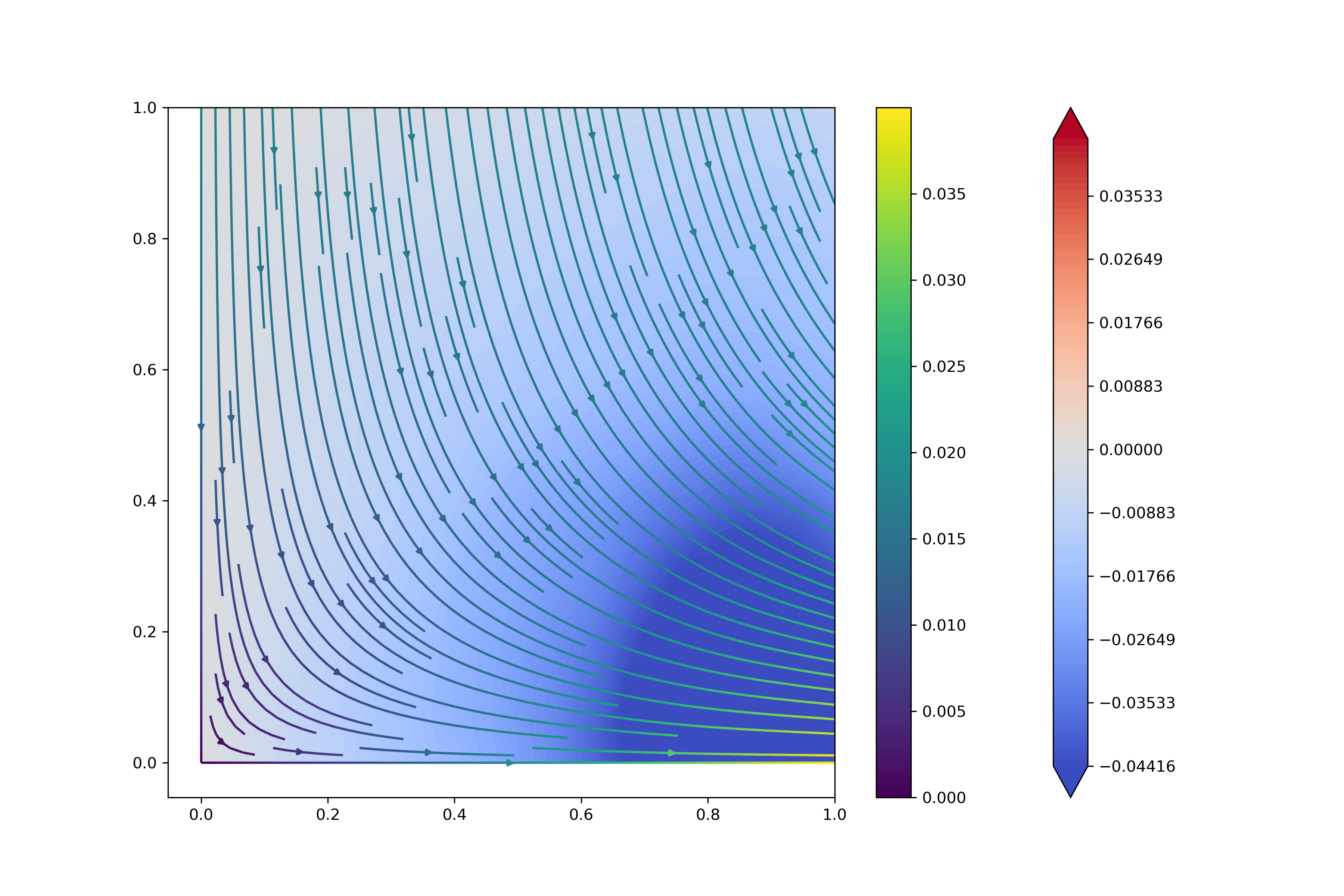}}}
    \qquad
    \subfloat[\centering $t=0.01$]{{\includegraphics[width=.3\linewidth]{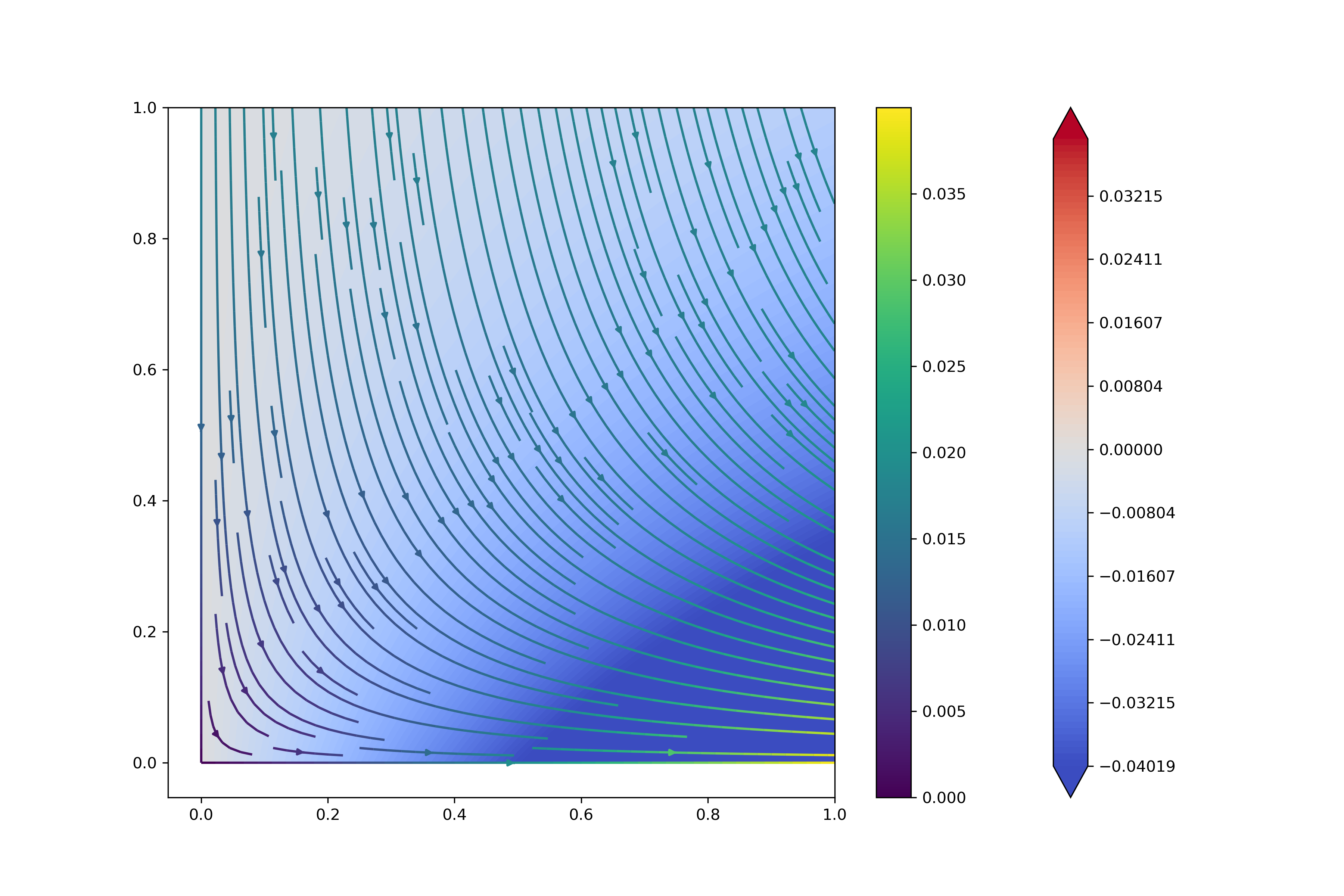} }}
    \subfloat[\centering $t=0.5$]{{\includegraphics[width=.3\linewidth]{Figures/true_flow.png} }}
    \subfloat[\centering $t=1.0$]{{\includegraphics[width=.3\linewidth]{Figures/true_flow.png} }}
    \qquad
    \subfloat[\centering $t=0.01$]{{\includegraphics[width=.3\linewidth]{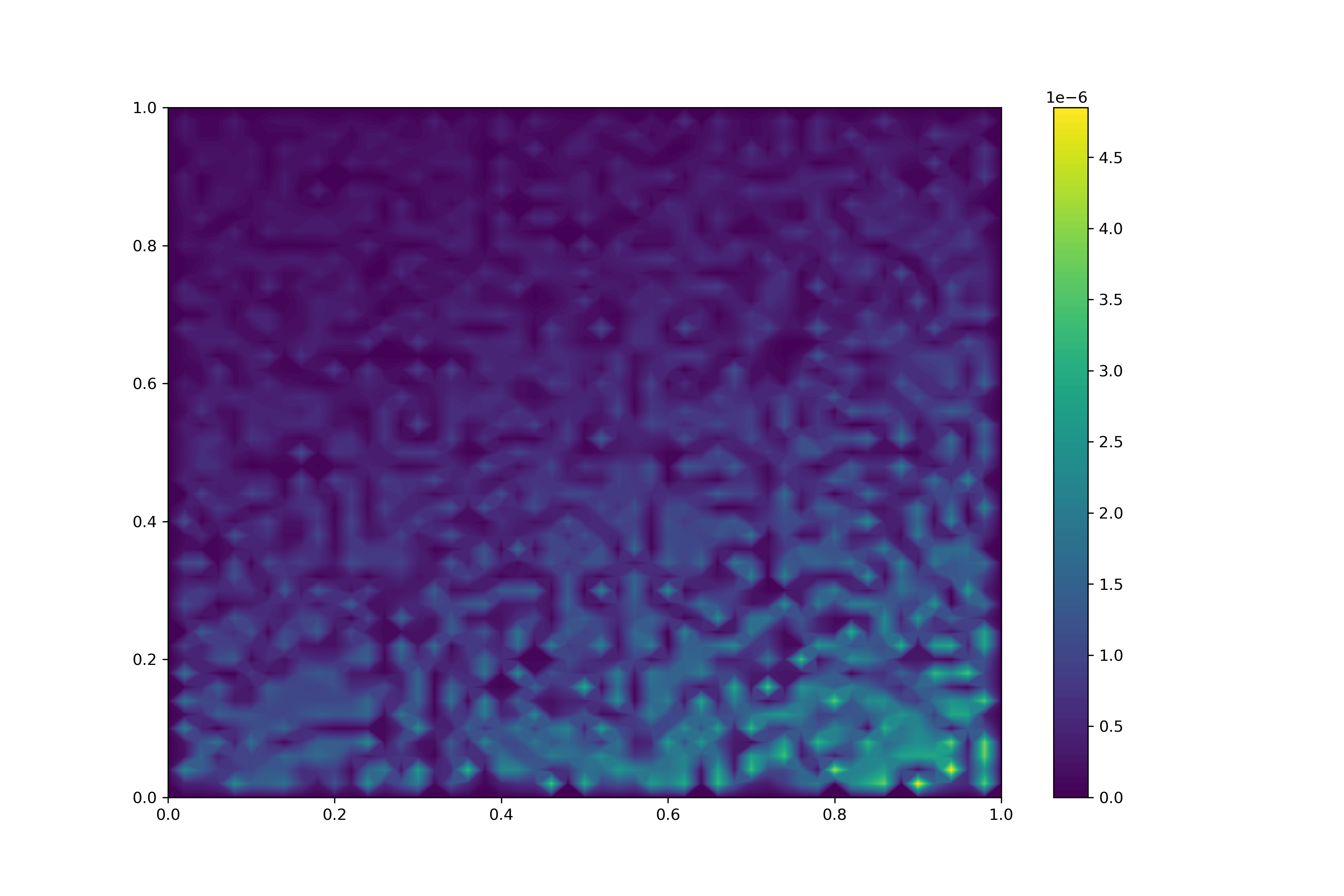} }}
    \subfloat[\centering $t=0.5$]{{\includegraphics[width=.3\linewidth]{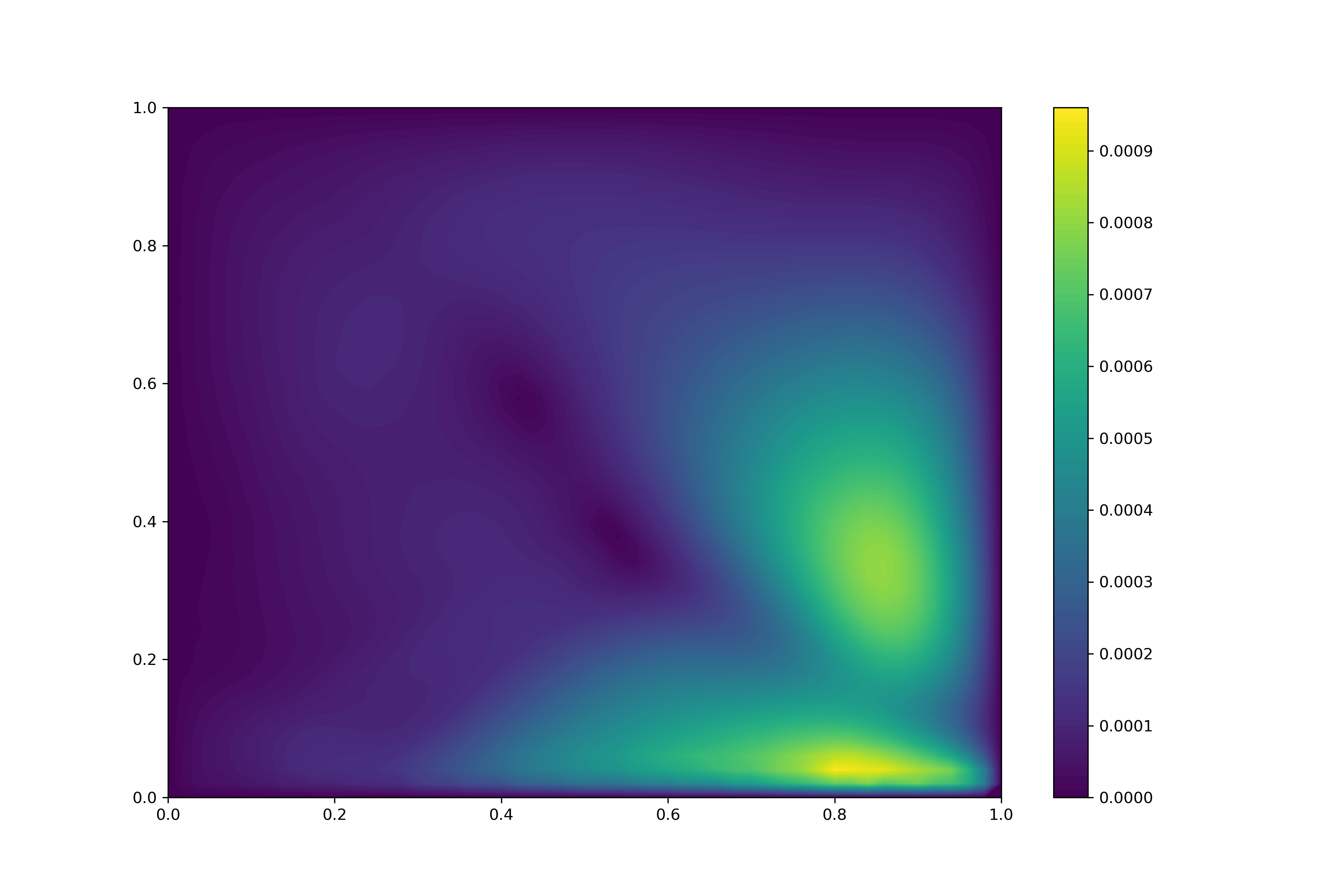} }}
    \subfloat[\centering $t=1.0$]{{\includegraphics[width=.3\linewidth]{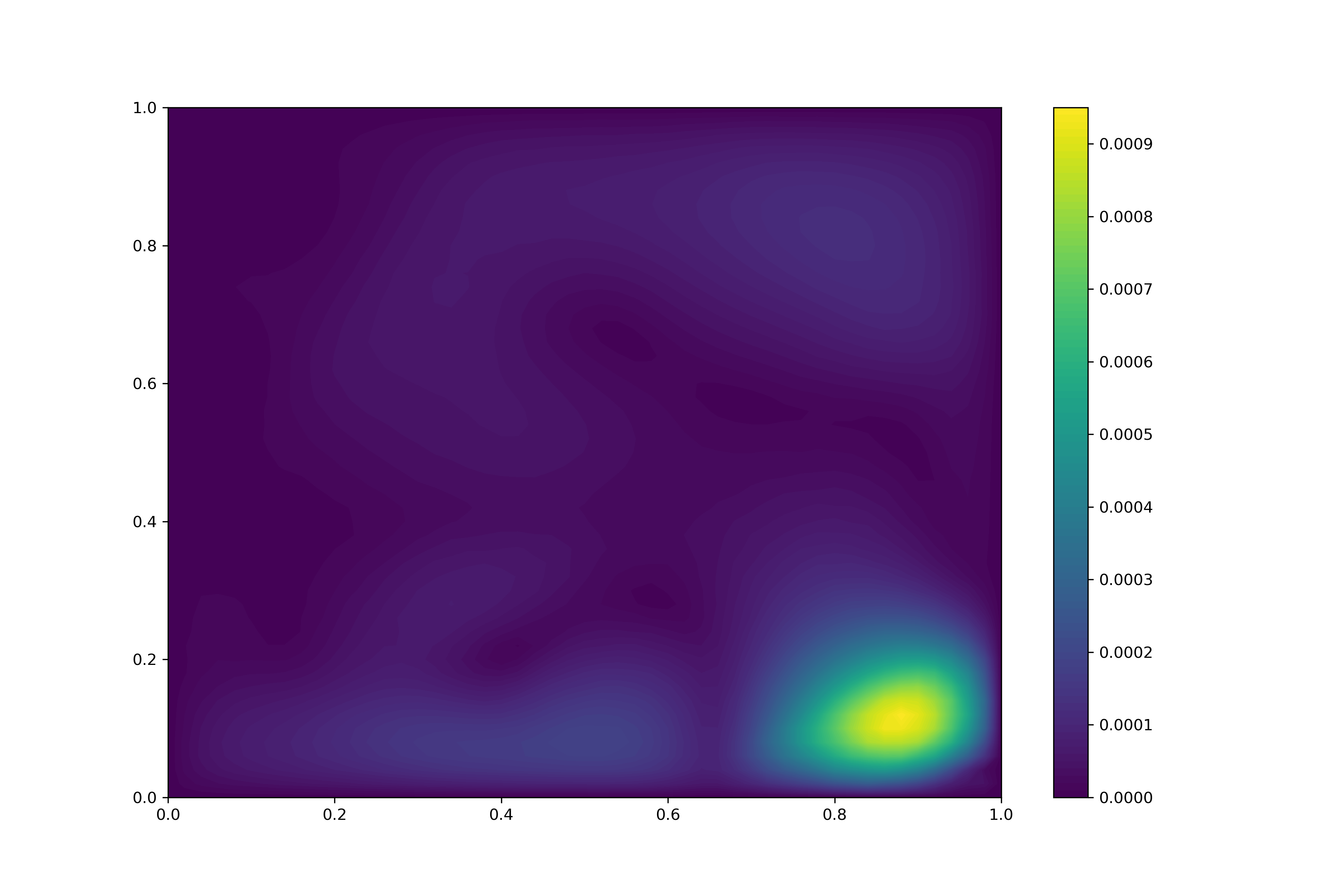} }}
    \caption{The fully-bounded flow \eqref{velocity_NE1}: the first row represents the approximation at different times $t$; the second row contains the true solution at the times; the third row represents the absolute l2-error of the approximation.}
    \label{fully_bounded_flow}
\end{figure}

\begin{figure}
    \centering
    \subfloat[\centering $t=0.01$]{{\includegraphics[width=.3\linewidth]{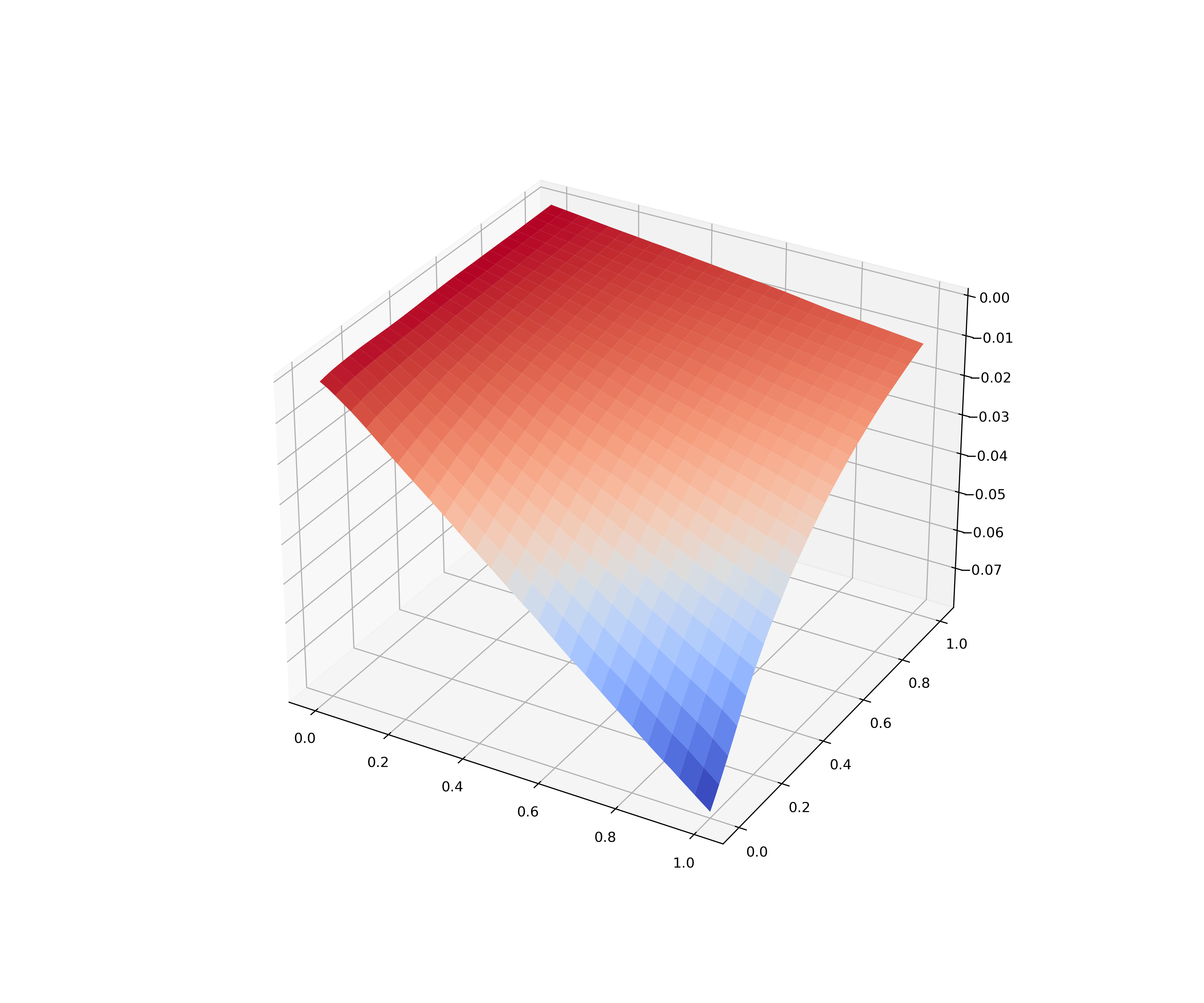} }}
    \subfloat[\centering $t=0.5$]{{\includegraphics[width=.3\linewidth]{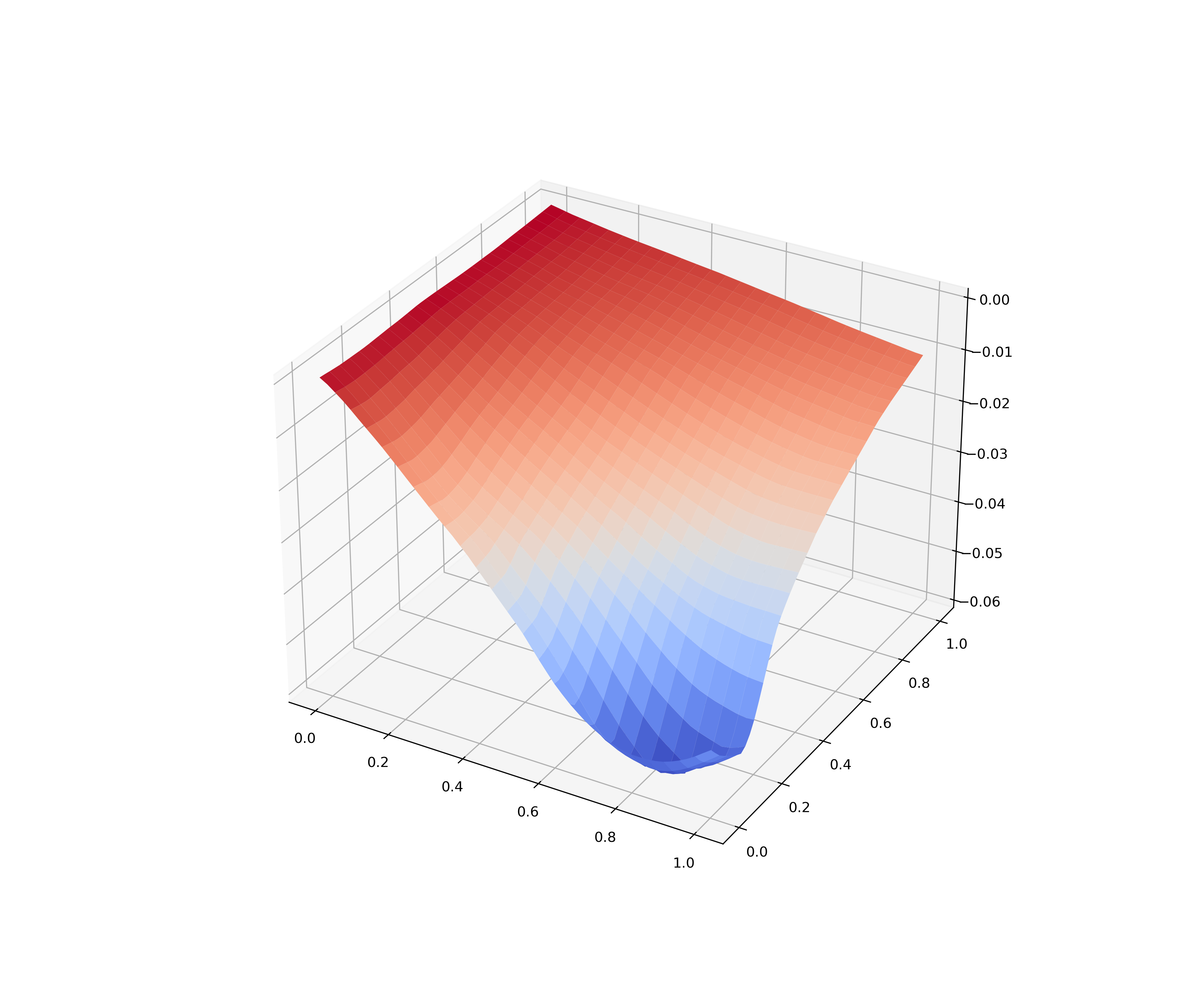}}}
    \subfloat[\centering $t=1.0$]{{\includegraphics[width=.3\linewidth]{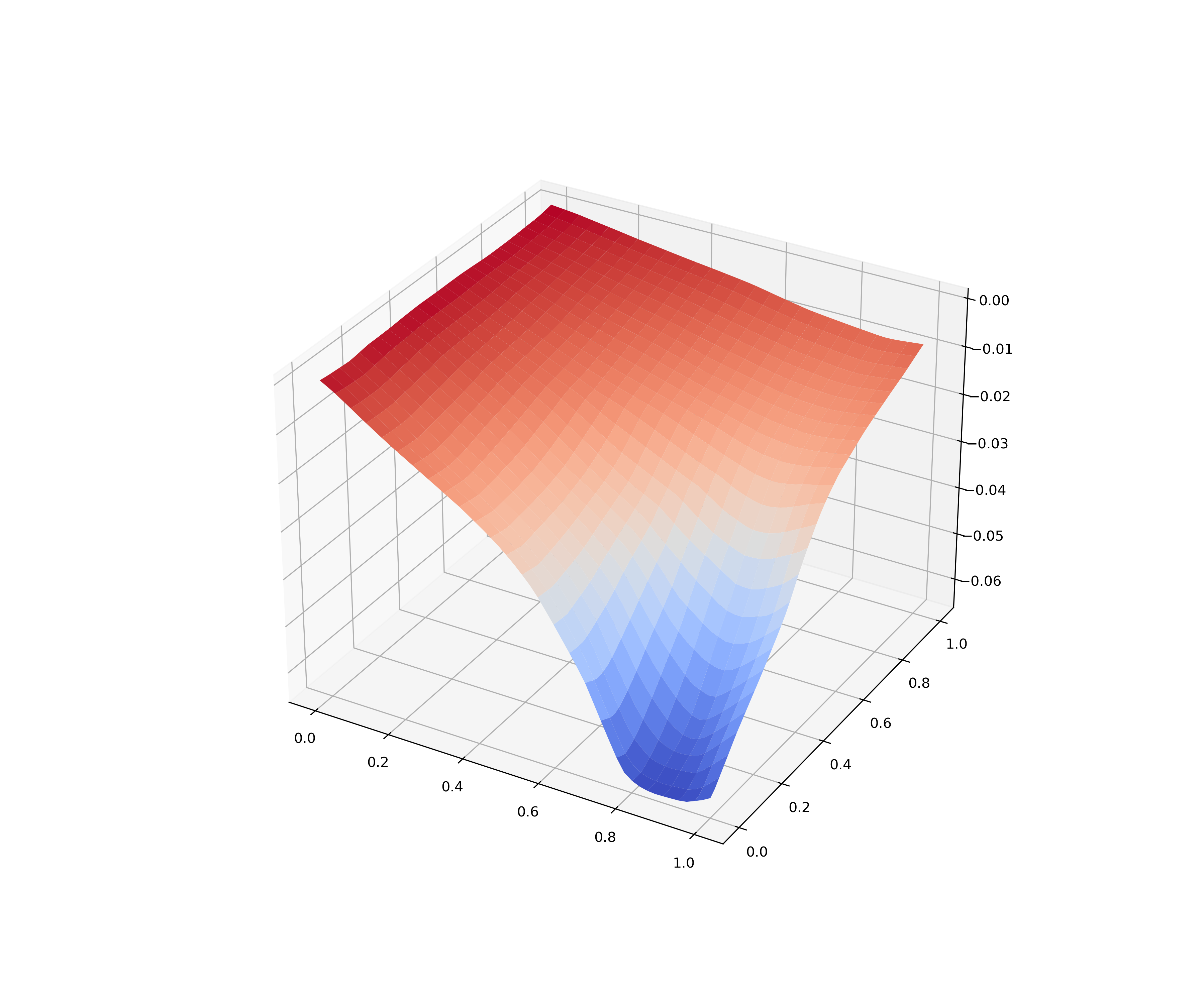}}}
    \qquad
    \subfloat[\centering $t=0.01$]{{\includegraphics[width=.3\linewidth]{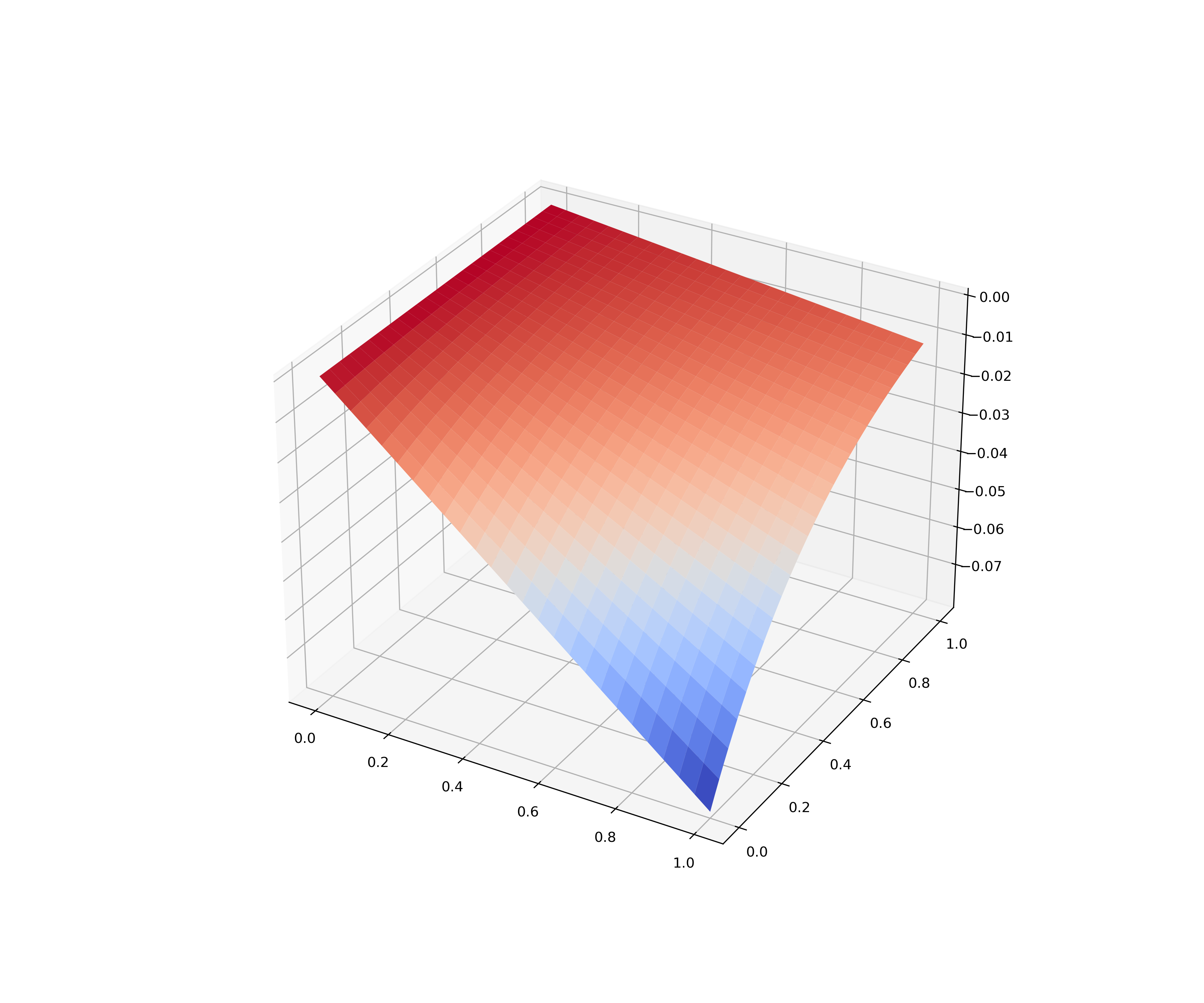} }}
    \subfloat[\centering $t=0.5$]{{\includegraphics[width=.3\linewidth]{Figures/3d_true_vorticity.png} }}
    \subfloat[\centering $t=1.0$]{{\includegraphics[width=.3\linewidth]{Figures/3d_true_vorticity.png} }}
    \caption{The fully-bounded flow \eqref{velocity_NE1}: the first row contains the vorticity learned at different times $t$; the second represents the true vorticity at the times.}
    \label{fully_bounded_flow_vorticity}
\end{figure}

    The simulation is conducted for $N_{\text{steps}} = 100$ iterations with time steps $h_{\text{time}}=0.01$ which gives the total time of $T = 1$, see Figure \ref{fully_bounded_flow} for the plots of the approximated and true flows for times $t = 0.01, 0.5$ and $1.0$. In these figures, the streamlines are colored by the velocity magnitude while the background is colored by the vorticity value. We also include a heatmap of the absolute l2-error of our approximated solution in Figure \ref{fully_bounded_flow}, and plots of the learned and true vorticity in Figure \ref{fully_bounded_flow_vorticity}.
    
    Our experiment showed that our NN-based algorithm produces velocity/vorticity approximations that are close to the true solution. Particularly, one can see that error shown in Figure \ref{fully_bounded_flow} is zero at the boundary which means that our algorithm enforces the correct inhomogeneous boundary conditions for the approximated velocity.

\begin{figure}
    \centering
    \includegraphics[width=8cm]{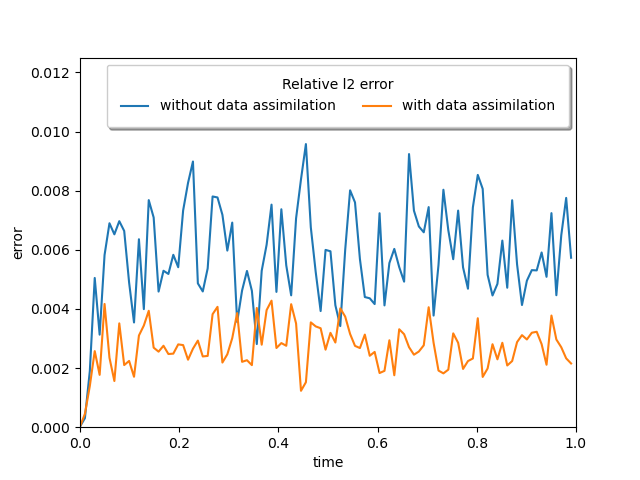}
    \caption{The relative l2 error with and without data assimilation as function of time $t$.}
    \label{rel_error_curves}
    \end{figure}

    Another advantage of the proposed numerical algorithm is also outlined in Section \ref{appendix - PINN} which allows for improving the quality of the results by adding additional penalisation for the approximation mismatching the measurement data. In order to demonstrate this in our experiment, we run the numerical scheme sampling at every timestep $t$ points $z_i, i = 1, \ldots, 2601$ independently and uniformly from $D$. We compute our "measurement data" as $\omega_i = \omega(z_i, t) + \sigma \varepsilon_i$ for $\varepsilon_i$ being independent standard normal and $\sigma = 5 \times 10^{-3}$, where $\omega$ is the true vorticity found from \eqref{velocity_NE1}. We use the values $\omega_i$ in the l2-loss \eqref{L2_loss_vorticity} and add this term to our loss function \eqref{discretized_loss}. The relative l2-errors of the approximations found with the original loss and the assimilated data versions are given in Figure \ref{rel_error_curves} which clearly demonstrates the improvement of adding data misfit penalisation.
    
    Some further details of the algorithm implementation and the model training are discussed in this section below.

    \textit{Numerical Experiment 2 [Taylor--Green vortex].} Consider $D$ to be the torus which is represented by the periodized box $[0, 1] \times [0, 1]$ and we assume that the boundary $\partial D = \varnothing$. We take the lattice $\Lambda := \{(i_1 h, i_2 h): 0 \leq i_1 \leq N, 0 \leq i_2 \leq N\}$ as above and choose the same parameters: the viscosity constant $\nu = 0.05$ and $N = 50$ implying $h = 0.02$. For this experiment, we take the Taylor-Green vortex which is given in our domain as
    
    \begin{equation}\label{velocity_NE2}
    u(x_1,x_2,t) = \left(-U_0 \sin(2\pi x_1) \cos(2\pi x_2) e^{-2\nu t}, U_0 \cos(2\pi x_1) \sin(2\pi x_2)e^{-2\nu t}\right)
    \end{equation}

    The initial vorticity is therefore taken as $\omega_0(x_1, x_2) = -U_0 4\pi \sin(2\pi x_1) \sin(2\pi x_2)$ with $U_0 = \frac{1}{2\pi}$. 

\begin{figure}
    \centering
    \subfloat[\centering $t=0.01$]{{\includegraphics[width=.3\linewidth]{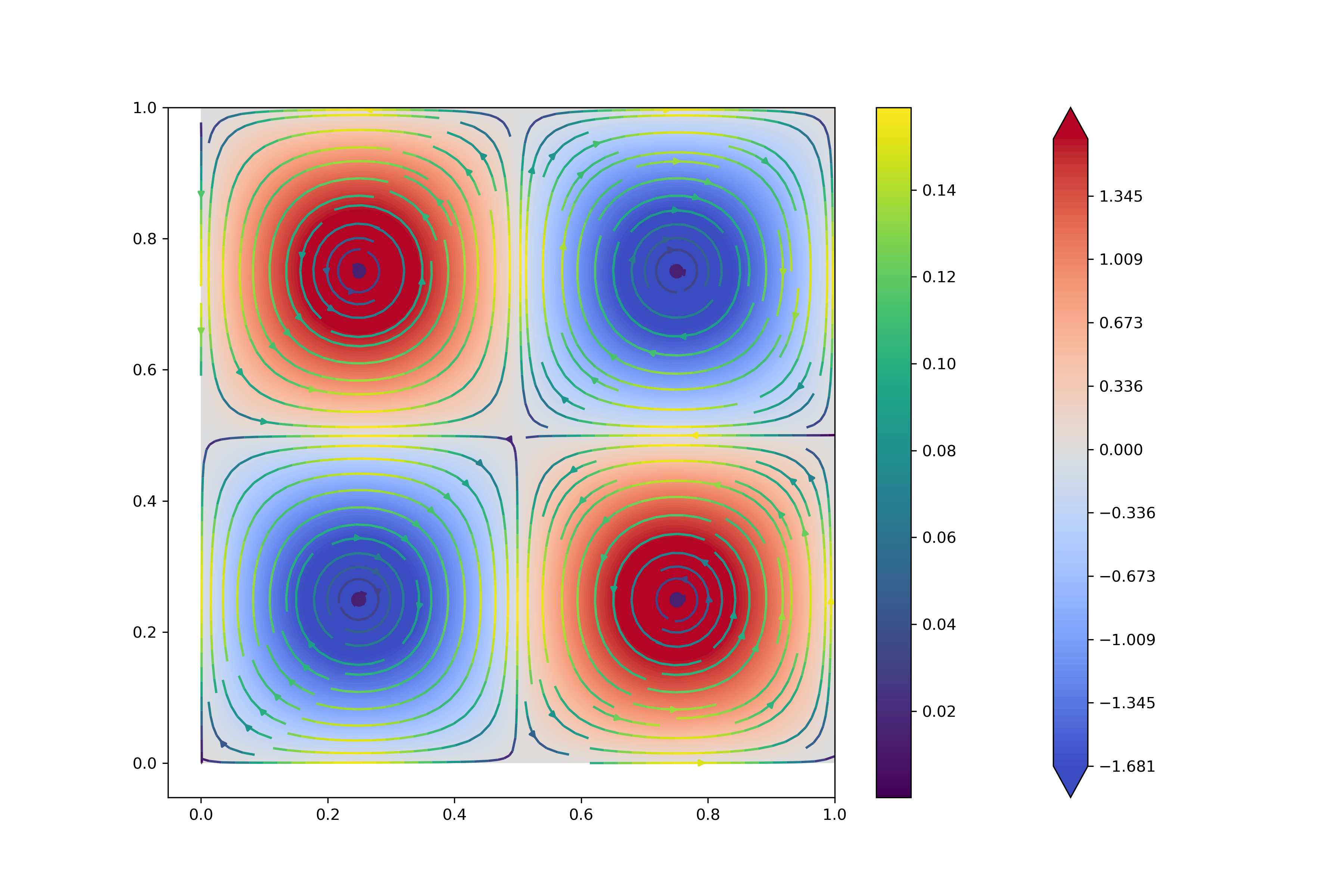} }}
    \subfloat[\centering $t=0.5$]{{\includegraphics[width=.3\linewidth]{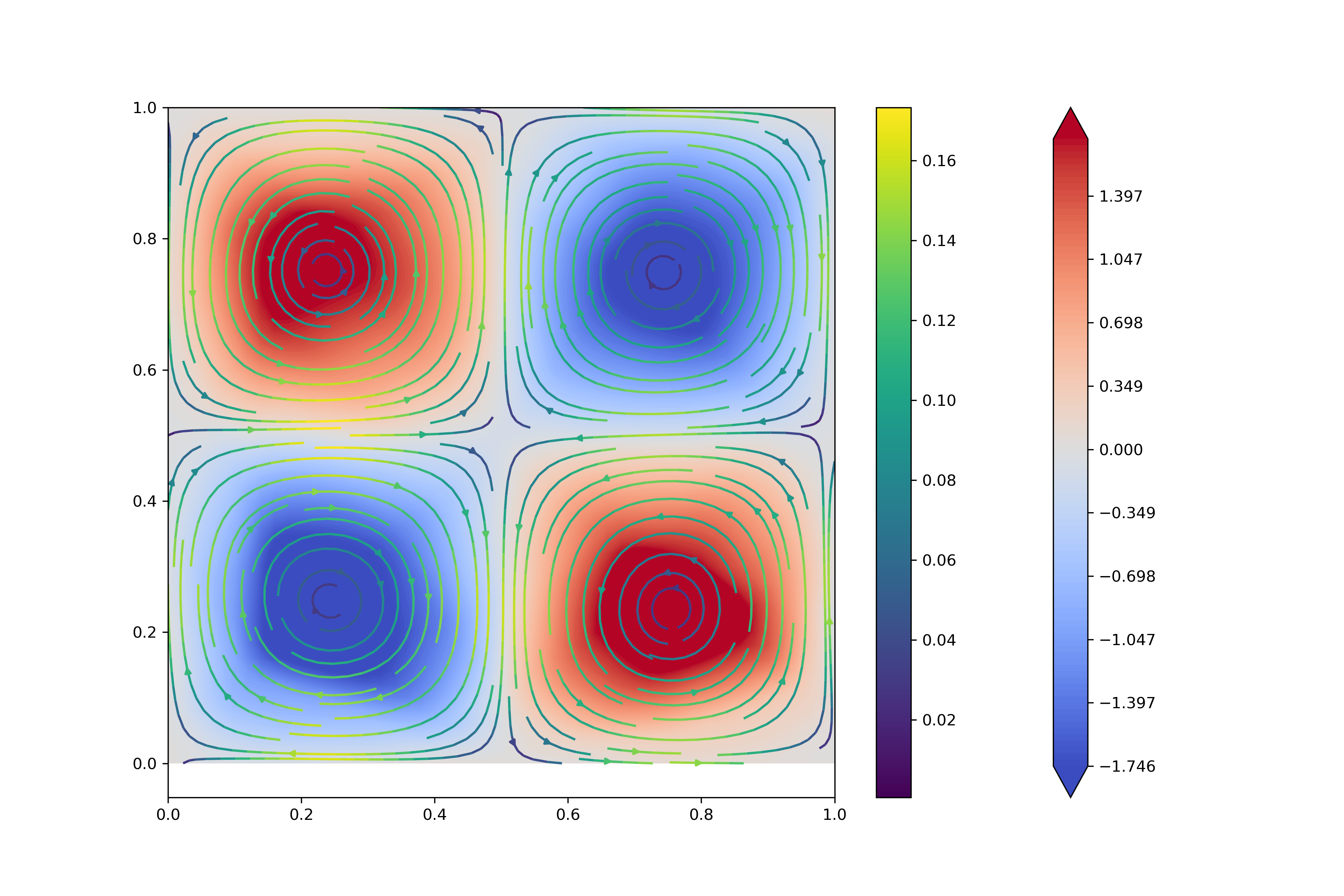}}}
    \subfloat[\centering $t=1.0$]{{\includegraphics[width=.3\linewidth]{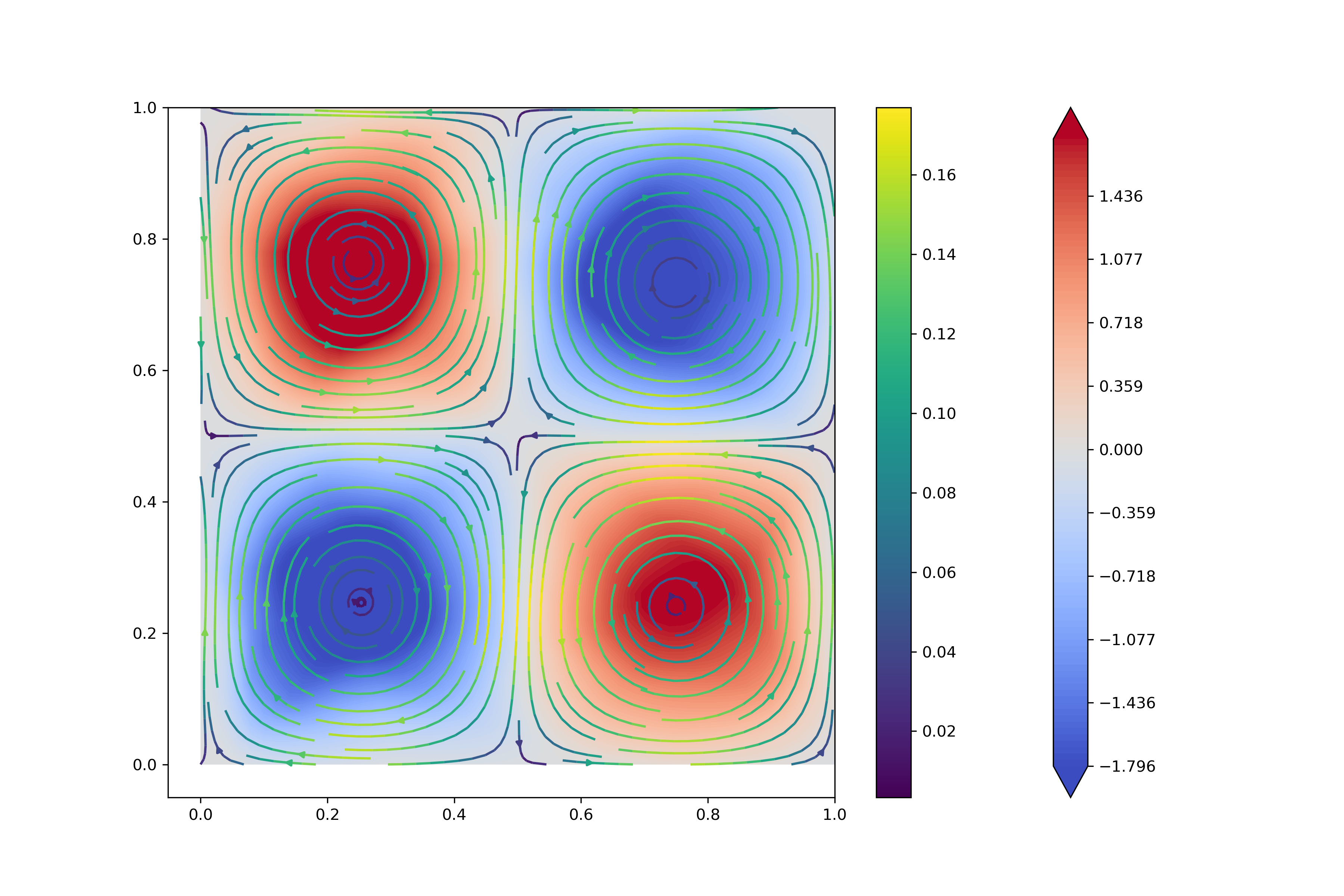}}}
    \qquad
    \subfloat[\centering $t=0.01$]{{\includegraphics[width=.3\linewidth]{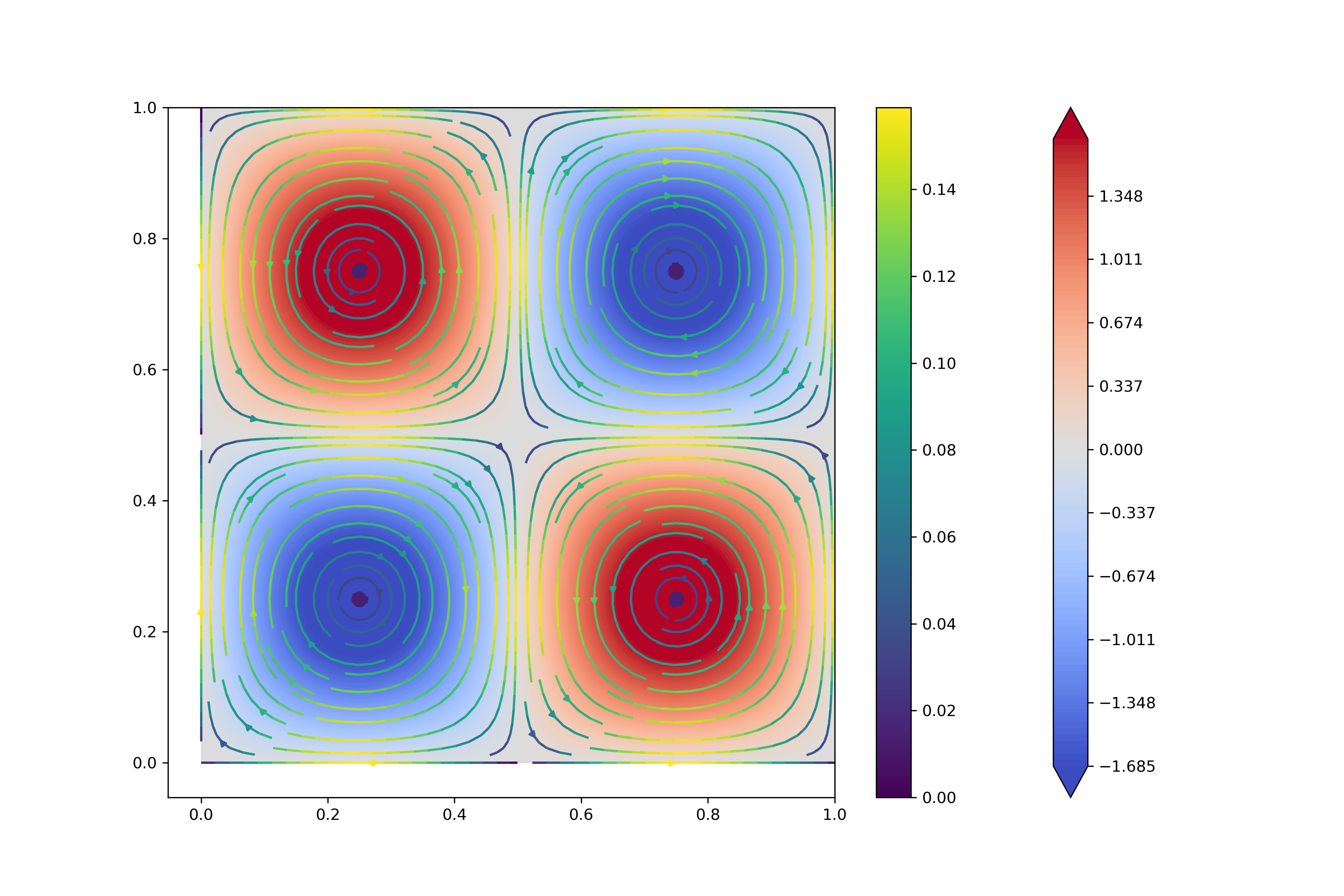} }}
    \subfloat[\centering $t=0.5$]{{\includegraphics[width=.3\linewidth]{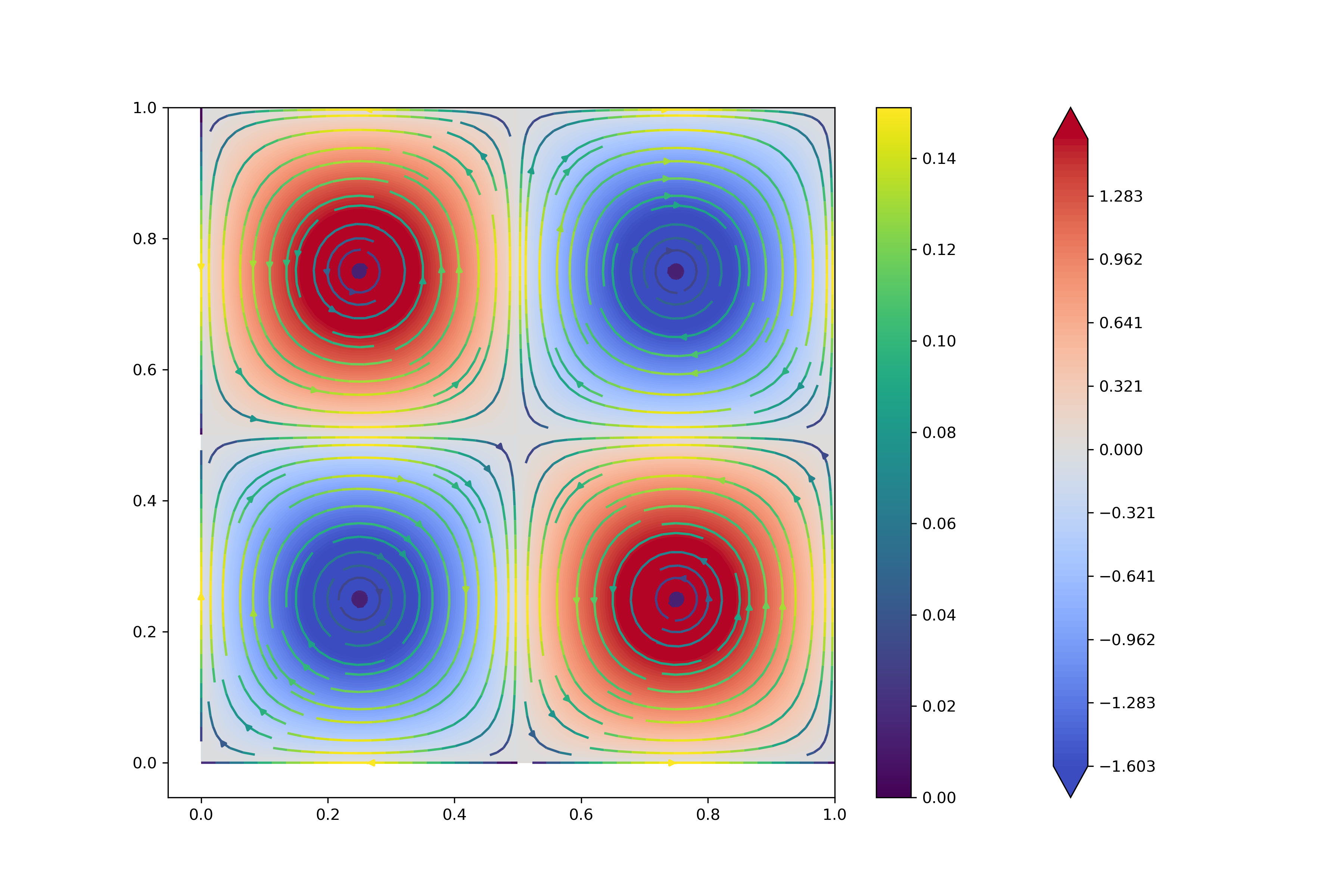} }}
    \subfloat[\centering $t=1.0$]{{\includegraphics[width=.3\linewidth]{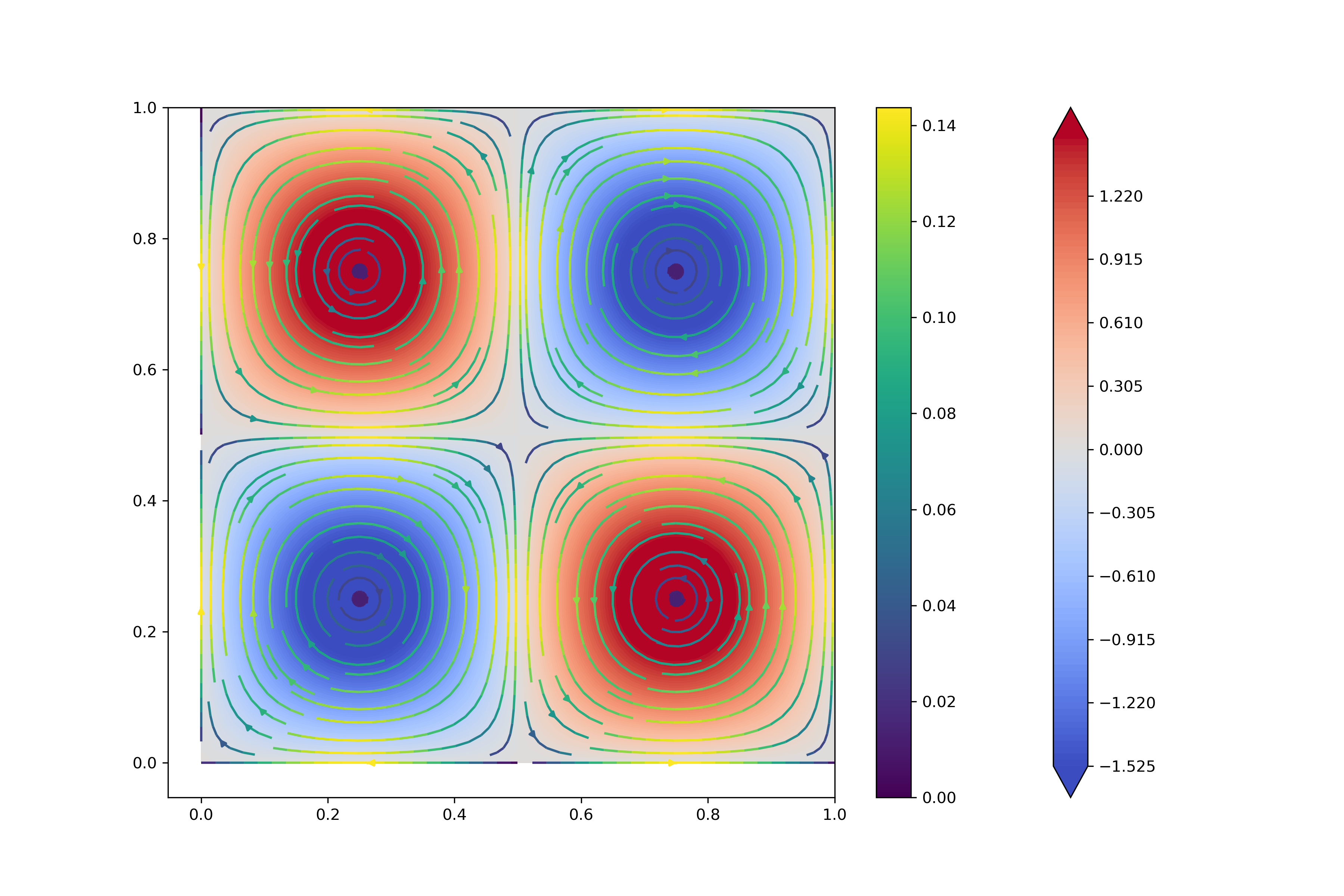} }}
    \qquad
    \subfloat[\centering $t=0.01$]{{\includegraphics[width=.3\linewidth]{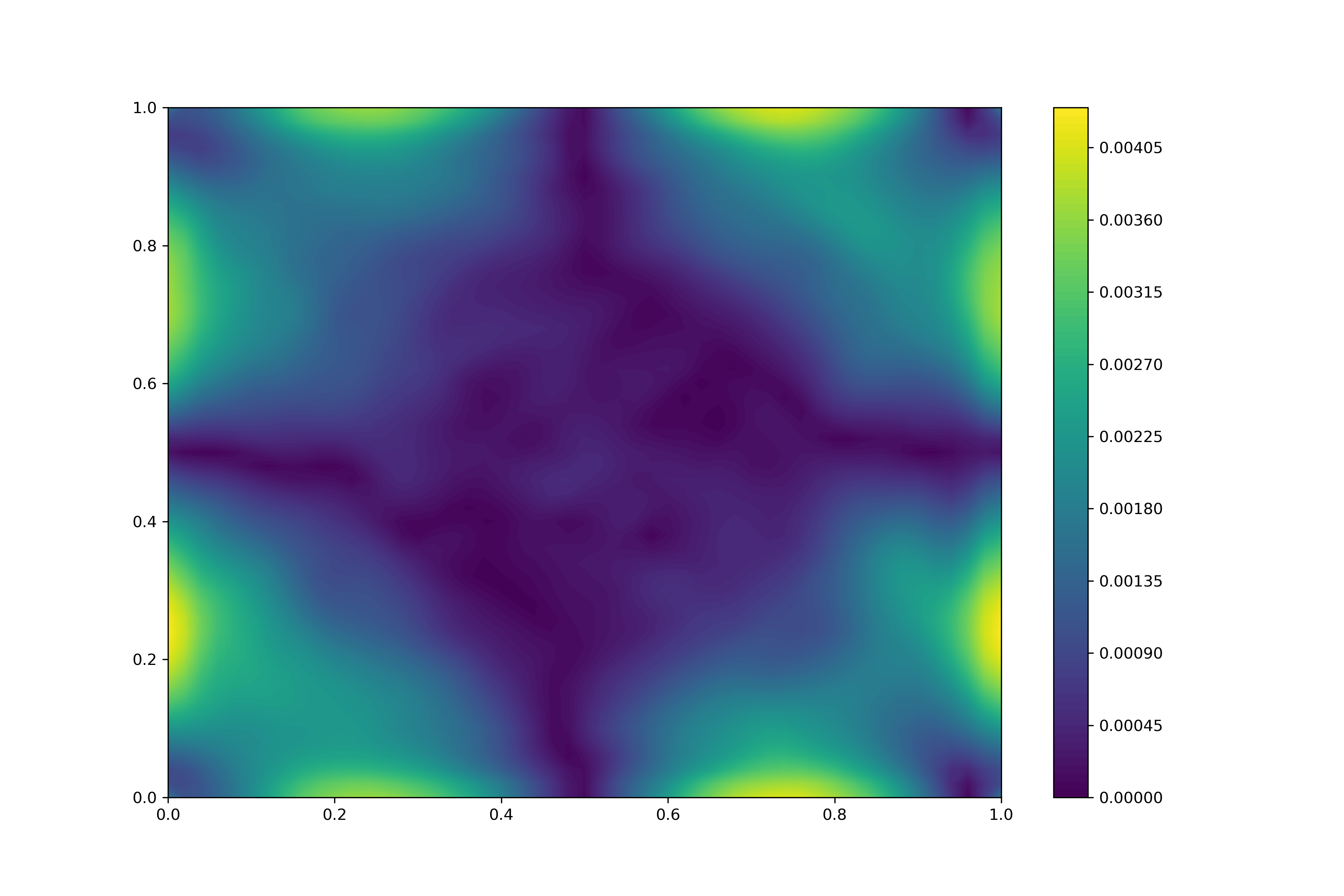} }}
    \subfloat[\centering $t=0.5$]{{\includegraphics[width=.3\linewidth]{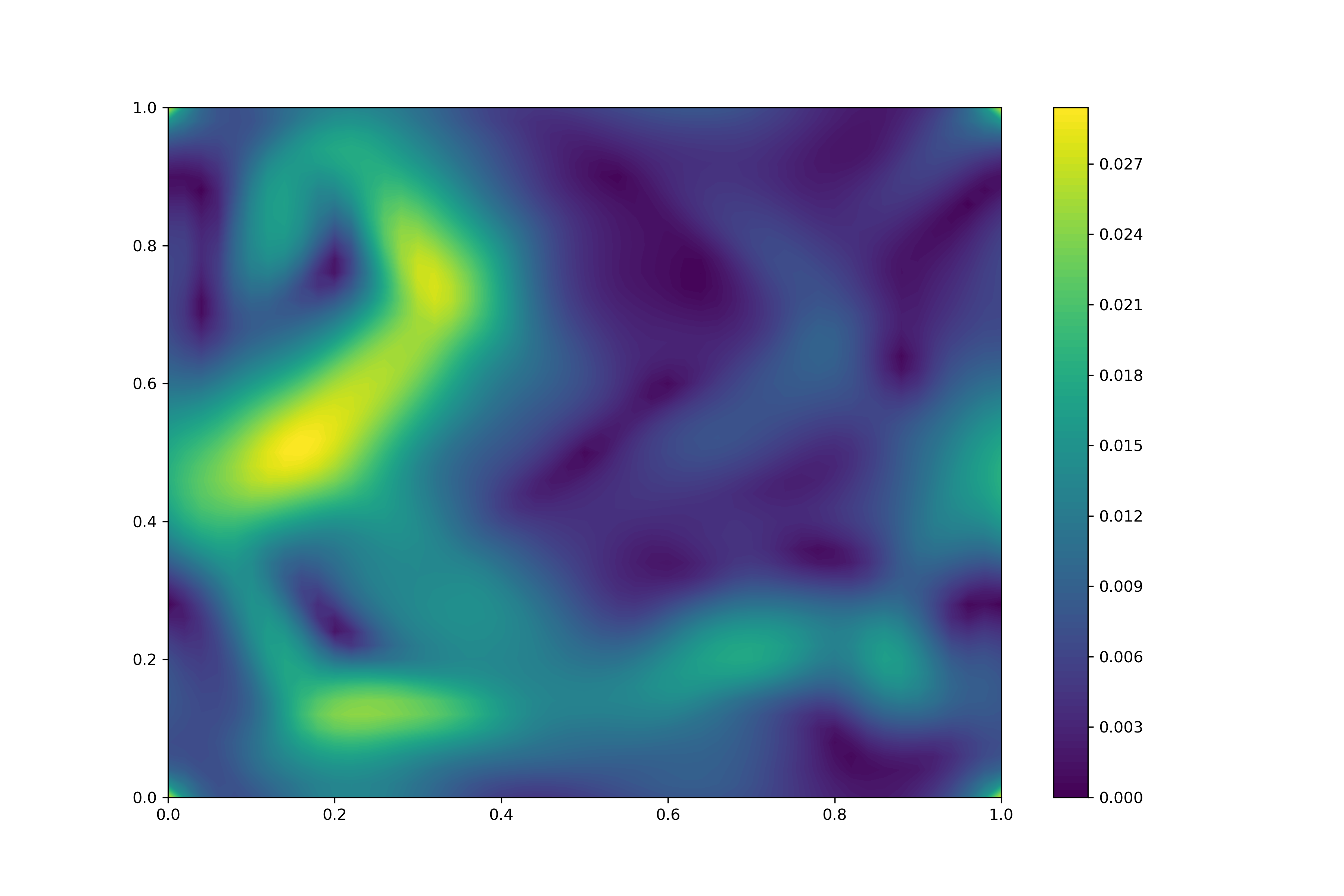} }}
    \subfloat[\centering $t=1.0$]{{\includegraphics[width=.3\linewidth]{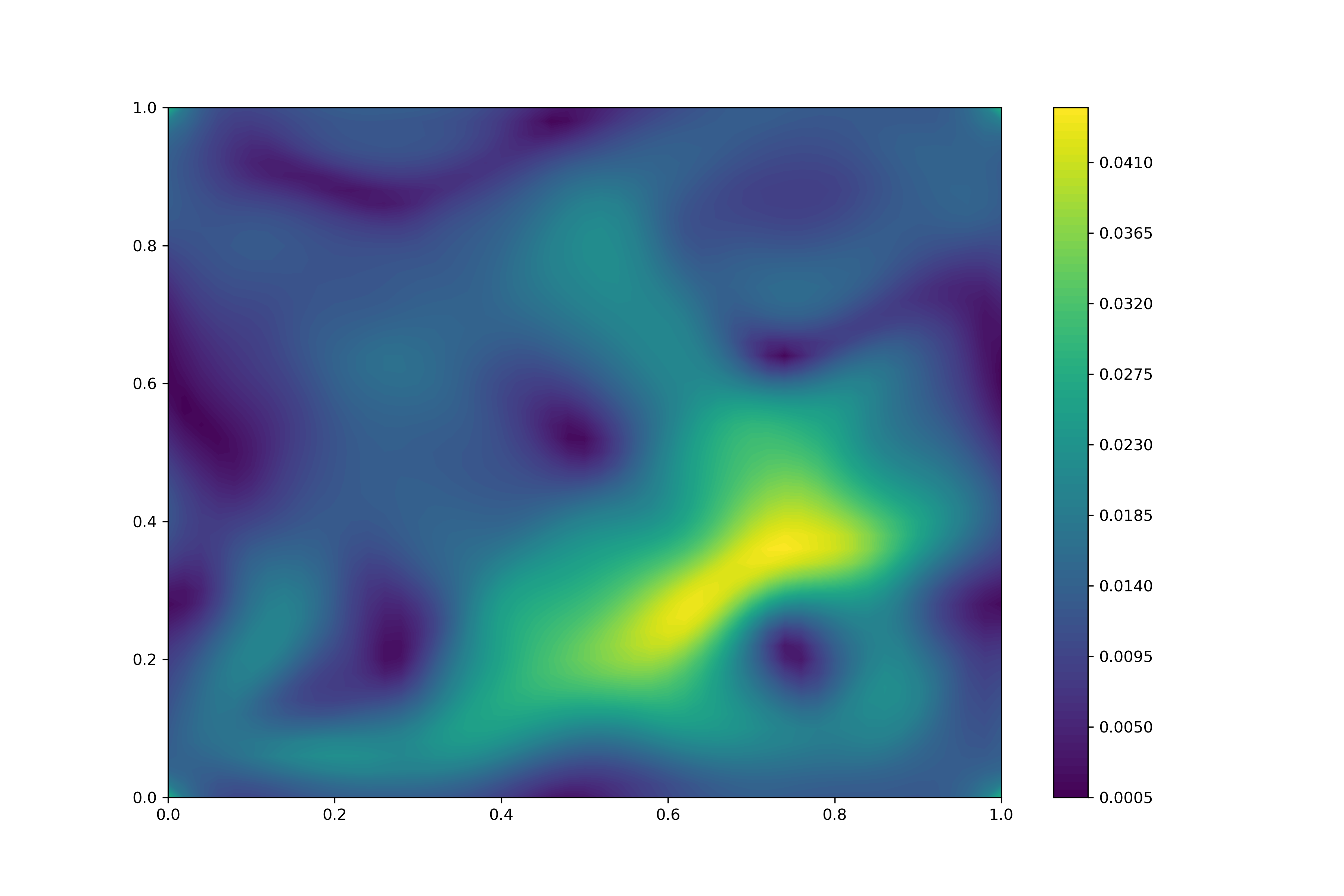} }}
    \caption{The fully-periodic flow \eqref{velocity_NE2} (Taylor--Green vortex): in the first and second row there are the approximation and the true solution at different times $t$; in the third row, the absolute l2-error of the approximation is printed.}
    \label{periodic_flow}
\end{figure}

    We again perform the simulation for $N_{steps} = 100$ iterations with the total time $T = 1$. The plots for the approximated flow are given in Figure \ref{periodic_flow} --- see the first row for the approximated solution at $t = 0.01, 0.5$ and $1$ against the true flow in the second row. We also print the heatmap for the absolute l2-error of the solution. Notice that the approximated solution matches the true Taylor--Green vortex closely, however, the average relative error in this experiment is $E_{[0,T]} \approx 0.1270$ compared to $E_{[0,T]} \approx 0.0142$ in the previous one. As the error at $t= 0.01$ (see Figure \ref{periodic_flow}) is concentrated along the boundary, this suggests that the error might be further improved by enforcing periodicity of the vorticity $\omega^{\theta}$ as discussed in Remark \ref{Period_Rem}. 

    \textit{Numerical Experiment 3 [Channel flow].}
    In this experiment we consider the periodic channel, which is represented by the box domain $D = [-H, H]\times[0, H]$ with boundary $\partial D = [-H, H]\times \{0\} \cup [-H, H]\times \{H\}$. We take similarly a uniform lattice $\Lambda = \{(i_1-N)h, i_2 h: 0 \leq i_1 \leq 2N, 0 \leq i_2 \leq N\}$ with $H = 6$ and $N = 100$ implying $h = 0.06$. The vorticity is initialised as
    \begin{equation}\label{period_channel_init_vort}
    \omega_0(x_1, x_2) = W_0 \cos(\frac{\pi}{3} x_2)
    \end{equation}
    with $W_0 = 5$; the vorticity constant $\nu = 0.01$ and the boundary cutoff $\varepsilon = 0.3$.

\begin{figure}
    \centering
    \subfloat[\centering $t=0.01$]{{\includegraphics[width=.45\linewidth]{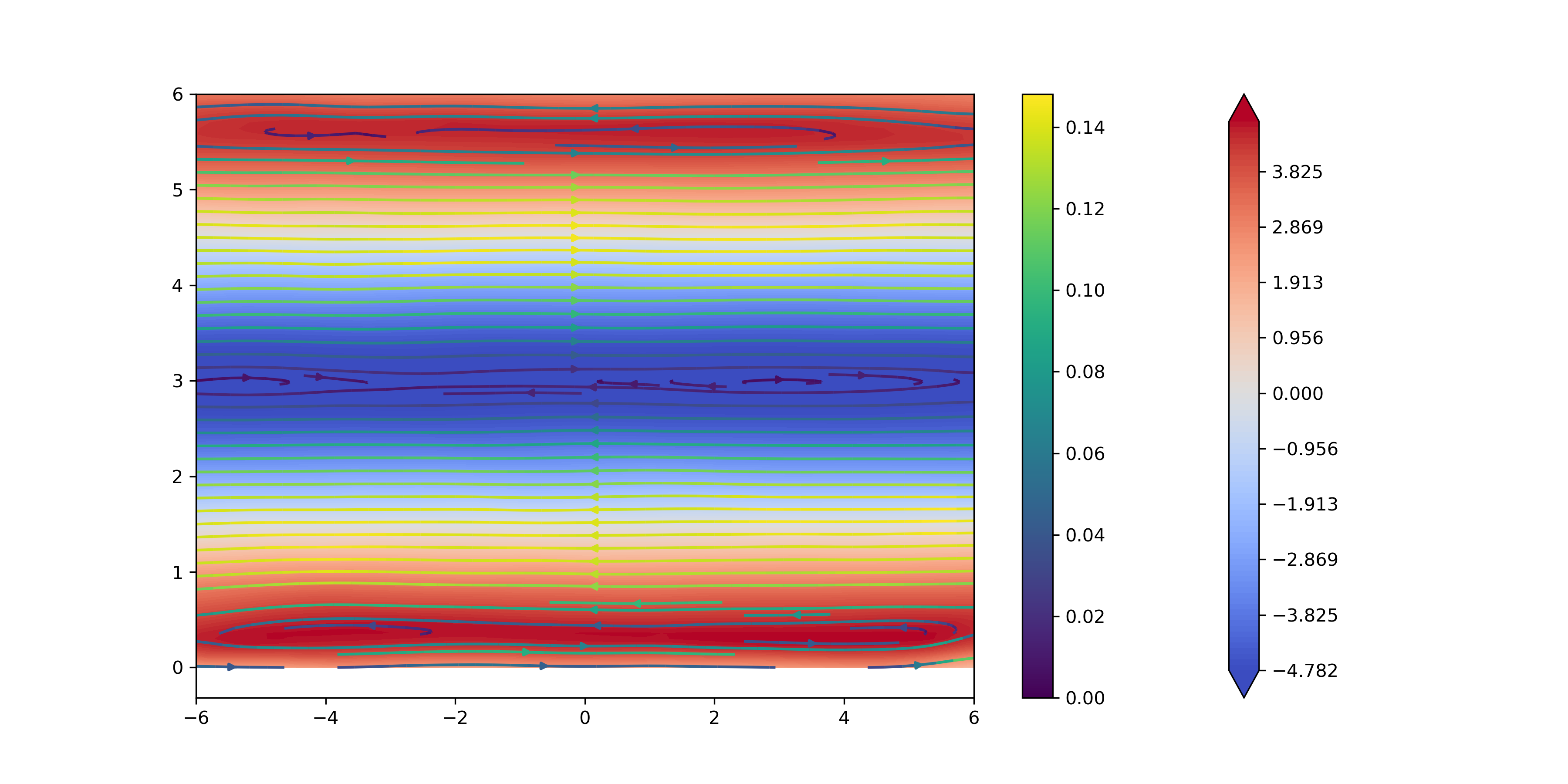} }}
    \subfloat[\centering $t=0.33$]{{\includegraphics[width=.45\linewidth]{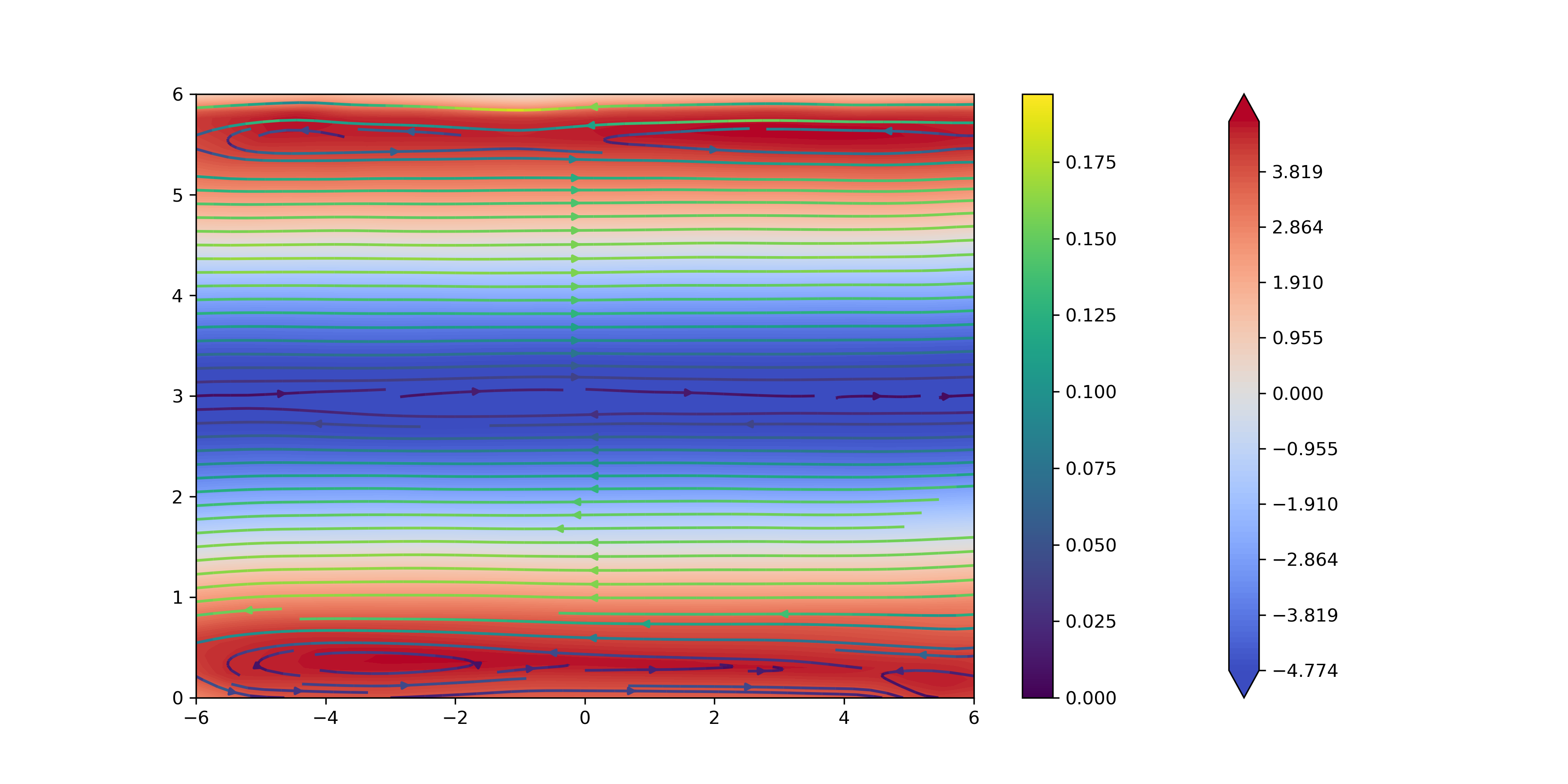}}}
    \qquad
    \subfloat[\centering $t=0.67$]{{\includegraphics[width=.45\linewidth]{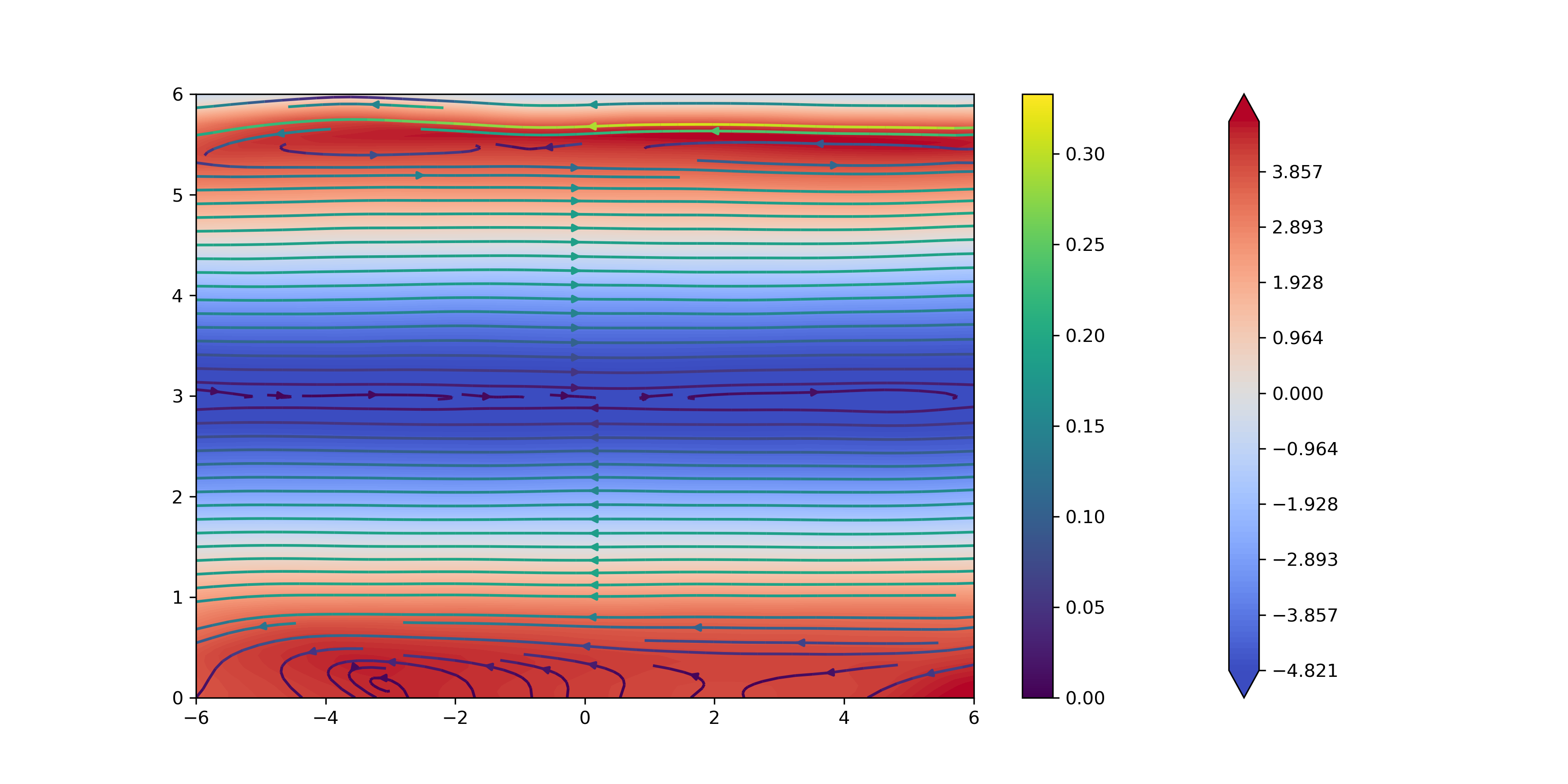} }}
    \subfloat[\centering $t=1.0$]{{\includegraphics[width=.45\linewidth]{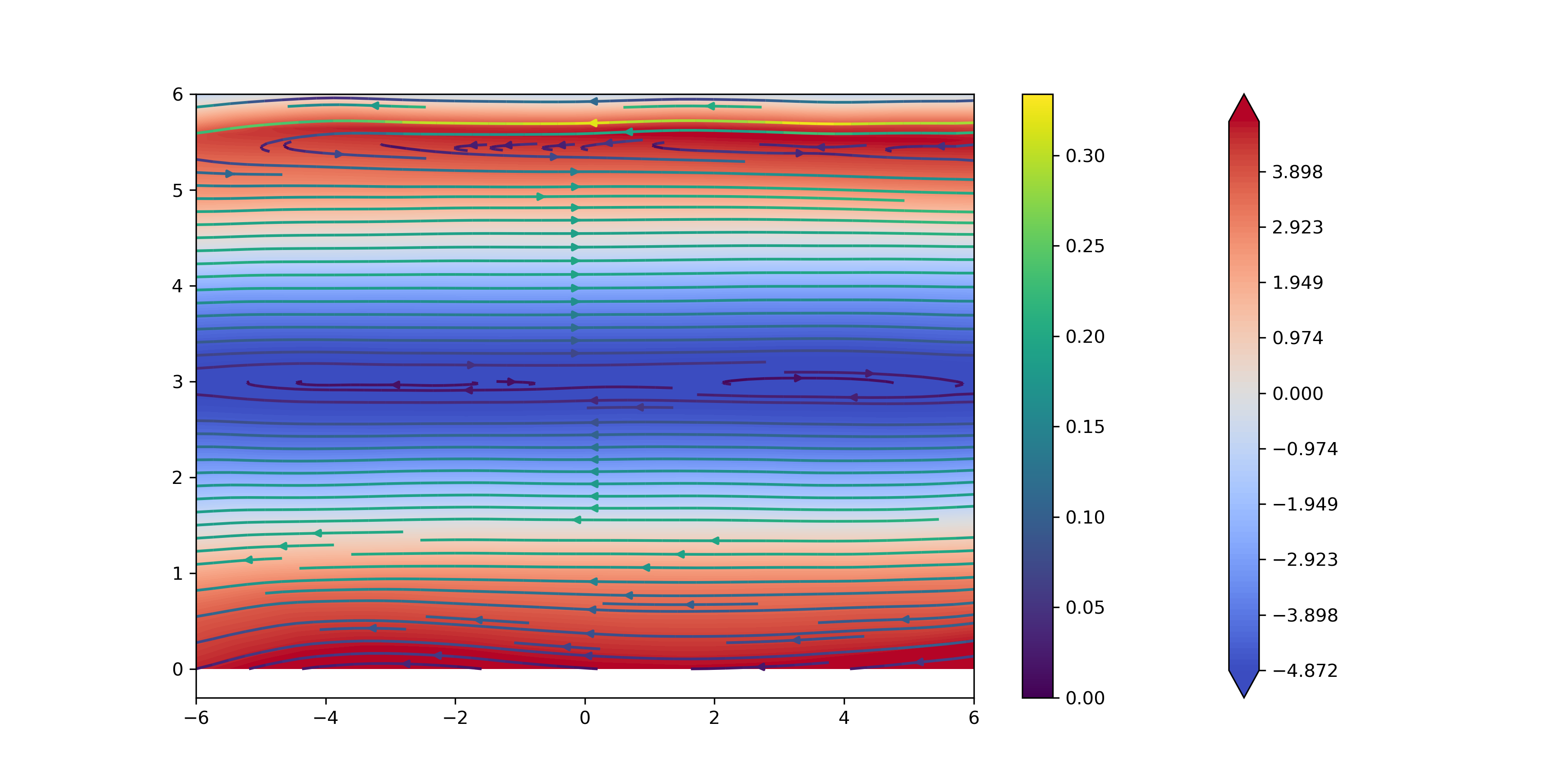} }}
    \caption{The periodic channel \eqref{period_channel_init_vort}: the approximated flow for different times $t$.}
    \label{channel_flow}
\end{figure}

    The results of this simulation are presented in Figure \ref{channel_flow} which includes the approximated streamlines and the vorticity of the flow for $t = 0.01, 0.33, 0.67$ and $1$.
    
    \textit{Algorithm implementation details.} Our implementation of the proposed algorithm is realized in Pytorch \cite{Pytorch}. At every iteration of the algorithm, we train a NN which represents the vorticity $\omega^{\theta}$. The architecture used in our simulations is a fully connected feedforward NN with 4 hidden layers of dimension 512, i.e. with the structure $(2, 512, 512, 512, 512, 1)$. We use the ReLU activation function after all hidden layers. 
    
\begin{figure}
    \centering
    \subfloat[\centering $t=0.5$]{{\includegraphics[width=.45\linewidth]{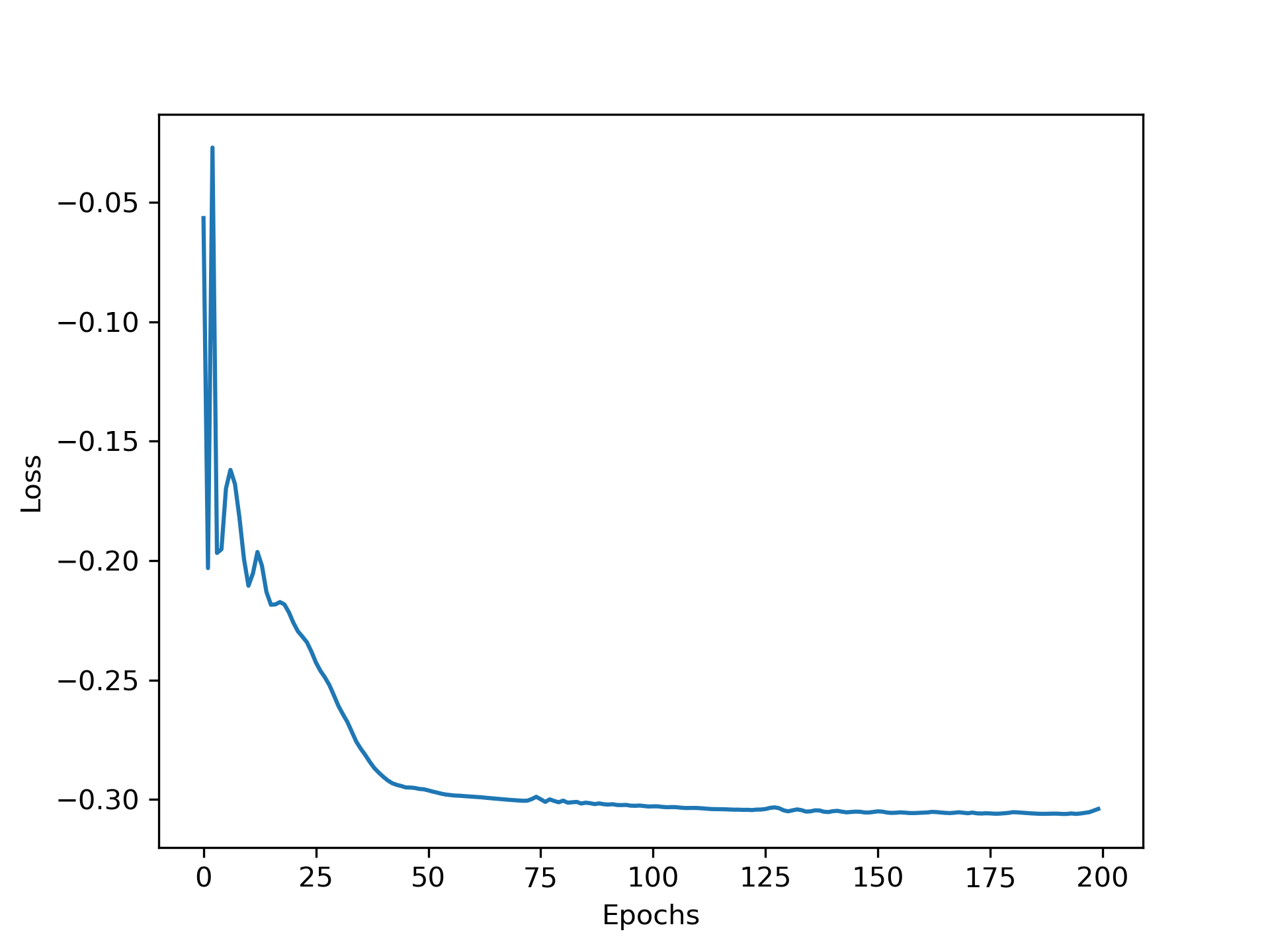} }}
    \subfloat[\centering $t=1.0$]{{\includegraphics[width=.45\linewidth]{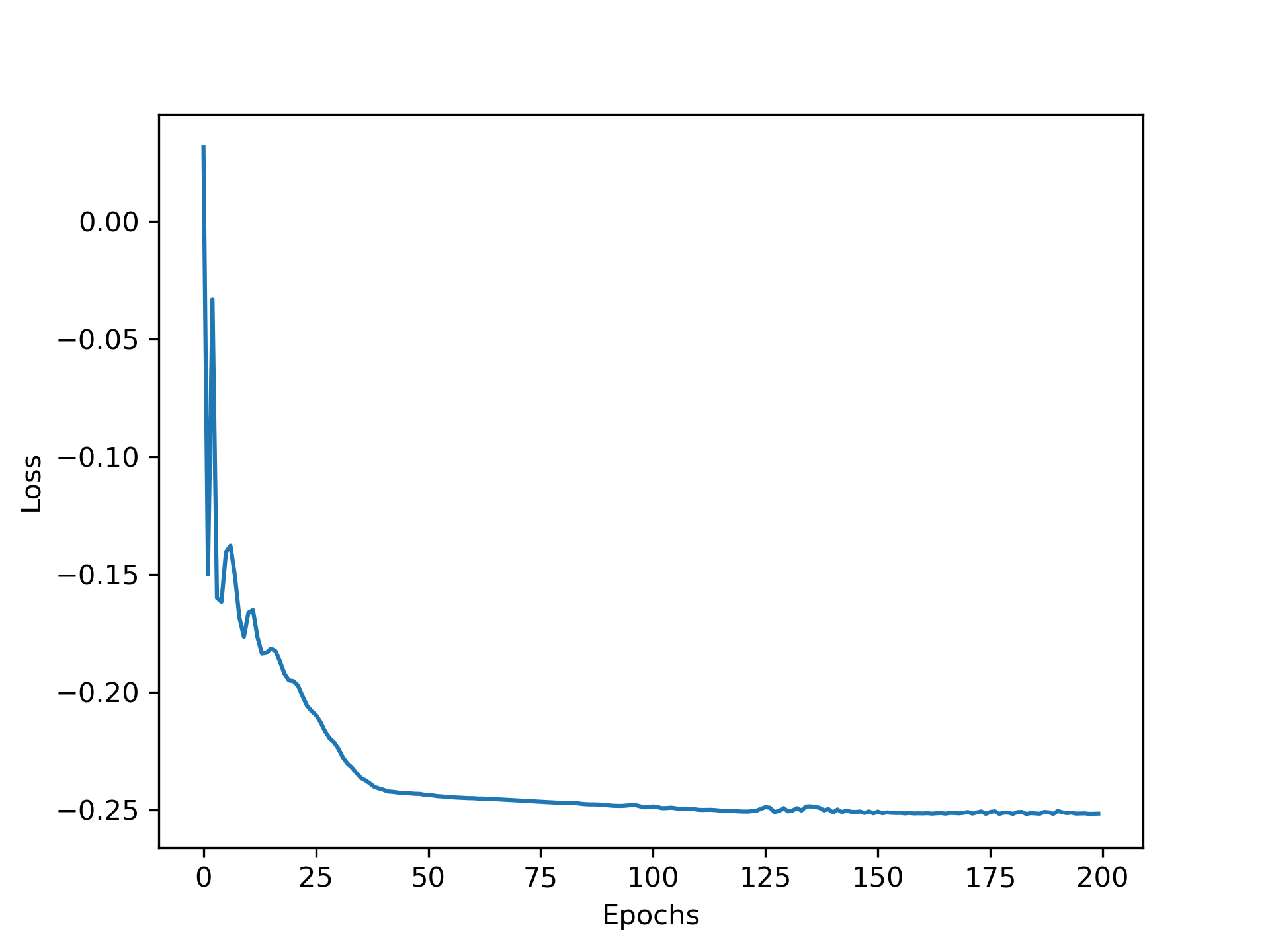}}}
    \caption{The loss function at different times $t$.}
    \label{loss_function_figure}
    \end{figure}

    At every step, we initialize a NN with the described architecture. For the initial weights, we use the default Pytorch initialization, i.e. uniform weights, and the optimization of the parameters is done with Adam \cite{KingmaBa 2015}. For each iteration, the model is trained for 500 epochs. However, we observe that in general the loss function values are stabilized before 100 epochs, see Figure \ref{loss_function_figure} for the loss function values for the first 200 epochs for the wall-bounded flow from Experiment 1. 
    
    % For training the vorticity from the loss function \eqref{discretized_loss}, we scale the values of the term $\Omega$ to be within the range $[0,1]$. After the training procedure, we descale the values of the model $\omega^{\theta}$ back into the original range, see Figure \ref{3d-vorticity_plot} for an example of the learned vorticity at time $t=1$ after descaling.

    To solve the Poisson equation, we use a Finite Element Method (FEM) for bounded domains (with Dirichlet or periodic boundary conditions). While a particular strength of our algorithm is that the Poisson solver can be used as a blackbox, allowing to employ the vast libraries of FEM software/toolboxes available, we instead used some simple custom made solvers working with uniform meshes on two-dimensional squares. In particular we did not consider more advanced algorithmic concepts such as adaptive mesh refinements.
    
    \begin{Rmk}
        While the uniform meshing we used can certainly be inefficient and lead to suboptimal usage of computational resources by the FEM, it should be noted that it allows for rapid evaluation of the interpolation of the computed gridbased values, which is import for fast simulation of the random vortex dynamics \eqref{vortex dynamics - definition}, and thus for training the network $\omega^{\theta}$.
    \end{Rmk}

    % As was mentioned above, the architecture and the hyperparameters were chosen so that the resulting model can replicate with sufficient precision the vorticity behavior found with the Monte-Carlo algorithm from \cite{CherepanovLiuQian} for the half-plane case. 
    
    Also notice that as in our implementation we use only simple fully-connected NNs and initialize them at every iteration, one potential way to upgrade the algorithm is to improve the NN architecture. For instance, one can include the time dependence into the NN and try to incorporate the Markovian structure, e.g. using some attention mechanism.

% \begin{figure}
%     \centering
%     \subfloat[\centering $t=0$]{{\includegraphics[width=.45\linewidth]{Figures_half-plane/outer_flow0.png} }}
%     \subfloat[\centering $t=0.25$]{{\includegraphics[width=.45\linewidth]{Figures_half-plane/outer_flow25.png}}}
%     \qquad
%     \subfloat[\centering $t=0.5$]{{\includegraphics[width=.45\linewidth]{Figures_half-plane/outer_flow50.png} }}
%     \subfloat[\centering $t=1.0$]{{\includegraphics[width=.45\linewidth]{Figures_half-plane/outer_flow100.png}}}
%     \caption{The half-plane flow at different times $t$.}
%     \label{half-plane_flow}
% \end{figure}

    % \begin{figure}
    % \centering
    % \subfloat[\centering $t=0.5$]{{\includegraphics[width=.45\linewidth]{Figures/stoch_force(t=0.5).png} }}
    % \subfloat[\centering $t=1.0$]{{\includegraphics[width=.45\linewidth]{Figures/stoch_force(t=1.0).png}}}
    % \caption{The flow with stochastic forcing at different times $t$.}
    % \label{loss_function_figure}
    % \end{figure}

\section*{Acknowledgements/Funding}
Vladislav Cherepanov is fully supported by the EPSRC Centre for Doctoral Training in Mathematics of Random Systems: Analysis, Modelling and Simulation (EP/S023925/1).\\

Sebastian W. Ertel has been supported by Deutsche Forschungsgemeinschaft (DFG) - Project-ID 318763901 - SFB1294 and by Deutsche Forschungsgemeinschaft (DFG) - Project-ID 410208580 - IRTG2544 ('Stochastic Analysis in Interaction').

\newpage

\bibliography{references}

\end{document}